\documentclass[times,onecolumn,final]{elsarticle}
\usepackage{color, colortbl}

\usepackage{adjustbox}
\usepackage{framed,multirow}
\usepackage{amssymb}
\usepackage{amsmath,mathtools}
\usepackage{changes}
\usepackage{realboxes}
\usepackage{tikz,pgfplots,adjustbox}

\usepackage{amssymb}
\usepackage{latexsym}
\usepackage[final]{pdfpages}
\usepackage{soul}
\usepackage[normalem]{ulem}
\usepackage[figuresright]{rotating}
\usepackage{url}
\usepackage{xcolor}
\usepackage{etoolbox}
\usepackage{lscape}
\usepackage{threeparttable}

\usepackage{hyperref}

\usepackage{cancel}

\definecolor{newcolor}{rgb}{.8,.349,.1}

\usepackage{lipsum}
\usepackage{booktabs}
\usepackage{caption}
\usepackage{pdflscape}
\usepackage{longtable}
\usepackage{lineno}
\usepackage{ragged2e}

\begin{document}

\begin{frontmatter}
\title{AI-based Aortic Vessel Tree Segmentation for Cardiovascular Diseases Treatment: Status Quo}%

\author[1,2,3]{Yuan Jin}
\author[1,2]{Antonio Pepe}
\author[1,2,4]{Jianning Li}
\author[1,2]{Christina Gsaxner}
\author[5]{Fen-hua Zhao}
\author [4]{Kelsey L. Pomykala}
\author[4,6,7]{Jens Kleesiek}
\author[8,9]{Alejandro F. Frangi}
\author[1,2,4,6]{Jan Egger \corref{cor}}

\address[1]{Institute of Computer Graphics and Vision, Graz University of Technology, Inffeldgasse 16, 8010 Graz, Austria}
\address[2]{Computer Algorithms for Medicine Laboratory, Graz, Austria}
\address[3]{Research Center for Connected Healthcare Big Data, ZhejiangLab, Hangzhou, Zhejiang, 311121 China}
\address[4]{Institute for AI in Medicine (IKIM), University Medicine Essen (AöR), Girardetstraße 2, 45131 Essen, Germany}
\address[5]{Department of Radiology, Affiliated Dongyang Hospital of Wenzhou Medical University, Dongyang, Zhejiang, 322100 China}
\address[6]{Cancer Research Center Cologne Essen (CCCE), University Medicine Essen (AöR), Hufelandstraße 55, 45147 Essen, Germany}
\address[7]{German Cancer Consortium (DKTK), Partner Site Essen, Hufelandstraße 55, 45147 Essen, Germany}
\address[8]{Centre for Computational Imaging and Simulation Technologies in Biomedicine (CISTIB), School of Computing, University of Leeds, Leeds, UK}
\address[9]{Biomedical Imaging Department, Leeds Institute for Cardiovascular and Metabolic Medicine (LICAMM), School of Medicine, University of Leeds, Leeds, UK}
\cortext[cor]{Corresponding author: Jan Egger (egger@tugraz.at)}

\begin{abstract}
The aortic vessel tree is composed of the aorta and its branching arteries, and plays a key role in supplying the whole body with blood. Aortic diseases, like aneurysms or dissections, can lead to an aortic rupture, whose treatment with open surgery is highly risky. Therefore, patients commonly undergo drug treatment under constant monitoring, which requires regular inspections of the vessels through imaging. The standard imaging modality for diagnosis and monitoring is computed tomography (CT), which can provide a detailed picture of the aorta and its branching vessels if completed with a contrast agent, called CT angiography (CTA). Optimally, the whole aortic vessel tree geometry from consecutive CTAs is overlaid and compared. This allows not only detection of changes in the aorta, but also of its branches, caused by the primary pathology or newly developed. When performed manually, this reconstruction requires slice by slice contouring, which could easily take a whole day for a single aortic vessel tree, and is therefore not feasible in clinical practice. Automatic or semi-automatic vessel tree segmentation algorithms, however, can complete this task in a fraction of the manual execution time and run in parallel to the clinical routine of the clinicians. In this paper, we systematically review computing techniques for the automatic and semi-automatic segmentation of the aortic vessel tree. The review concludes with an in-depth discussion on how close these state-of-the-art approaches are to an application in clinical practice and how active this research field is, taking into account the number of publications, datasets and challenges.
\end{abstract}


\end{frontmatter}

\thispagestyle{plain}
\section{Introduction}\label{sec1}

\begin{figure}
    \centering
    \includegraphics[width=0.7\linewidth]{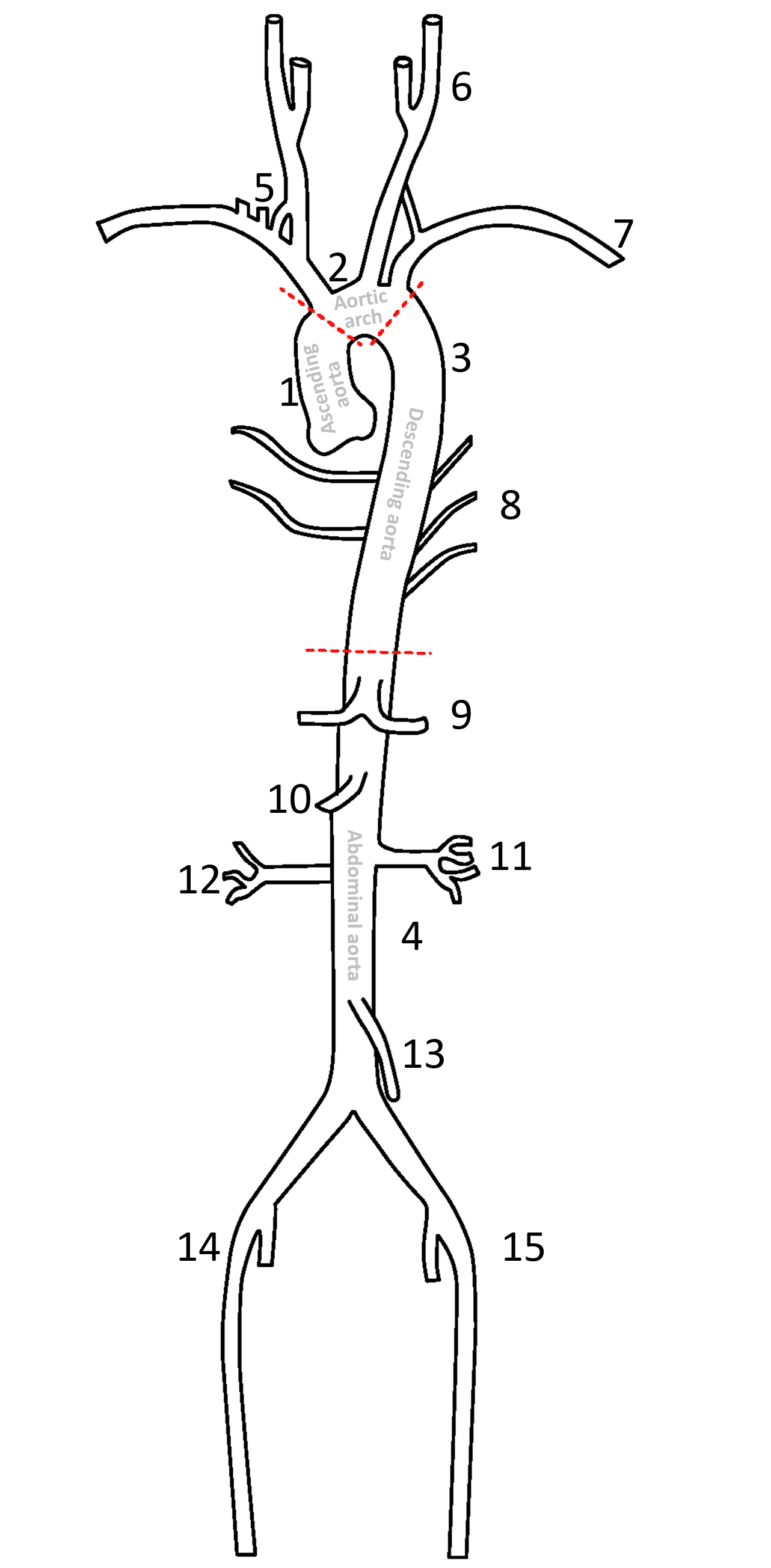}
    \caption{Aortic vessel tree and its main branches and labels (based on the aortic vessel tree and labels of~\cite{Du_2015}). Please see \autoref{tab:vesseltreelabels} for the corresponding labels. The figure also shows the main aortic segments -- separated by red dashed lines -- according to \cite{Erbel_2014}, originating from (the left ventricle) of the heart, extending down to the abdomen, where it splits into the common iliac arteries (labels 14 and 15): Ascending aorta, aortic arch, descending aorta and abdominal aorta. Image courtesy of Xiaoyan Zhou.}
    \label{fig:AorticVesselTree}
\end{figure}

The aorta is the largest artery in the human body. It is supplied by the left ventricle and consists of four main sections -- ascending aorta, aortic arch, descending thoracic aorta, and abdominal aorta. A schematic image is shown in \autoref{fig:AorticVesselTree} and \autoref{tab:vesseltreelabels}. The aorta delivers oxygen-rich blood from the heart to the whole body. Hence, diseases of the aorta, such as aneurysms, stenosis and dissections, can pose a significant threat to a patient's life. Non-invasive examinations of the aorta are therefore crucial for early detection and monitoring of these diseases~(\cite{Nienaber_2013,Tortora_2016}).

These non-invasive examinations include computed tomography and magnetic resonance imaging, with or without contrast agents~(\cite{Nienaber_2013,Tortora_2016,Nienaber_2016ADRev,Pepe_2020}). Segmentation of the whole aorta and all its branches, also named aortic vessel tree segmentation, is critically important during clinical examination of non-invasive vascular imaging~(~\cite{Mistelbauer_2016,Gamechi_2019,Hahn_2020}). Compared with common analysis of two-dimensional slices, the 3D reconstruction of the aortic vessel tree provides a more intuitive and comprehensive visualization. It can help radiologists visualize the whole aorta before examining the single cross-sectional images and help generate broad overviews that aid in CT comparison aimed at identifying disease progression~(\cite{Mistelbauer_2016}). In addition, the aortic vessel tree helps patients better understand their condition, which can support the communication between doctors and patients~(\cite{ritter2006real,lawonn2018survey}).

\begin{table}[ht]
  \centering
    \caption{Aortic vessel tree and its main branches (the vessel labels are used in~\autoref{fig:AorticVesselTree}).}
    \captionsetup{width=1\textwidth}
  \label{tab:vesseltreelabels}
  \begin{tabular}{c c}
                            \toprule
    Label&Name \\  \midrule
    1&Ascending aorta\\ 
    2&Aortic arch \\ 
    3&Descending thoracic aorta \\  
    4&Abdominal aorta \\  
    5&Brachiocephalic trunk\\
    6&Left common carotid \\
    7&Left subclavian artery\\
    8&Intercostal arteries \\
    9&Celiac trunk \\  
    10&Superior mesenteric artery\\
    11&Left renal artery\\
    12&Right renal artery\\
    13&Inferior mesenteric artery \\
    14&Right iliac artery\\
    15&Left iliac artery\\
    \bottomrule
  \end{tabular}
\end{table}

\begin{figure}
    \centering
    \includegraphics[width=1.0\linewidth]{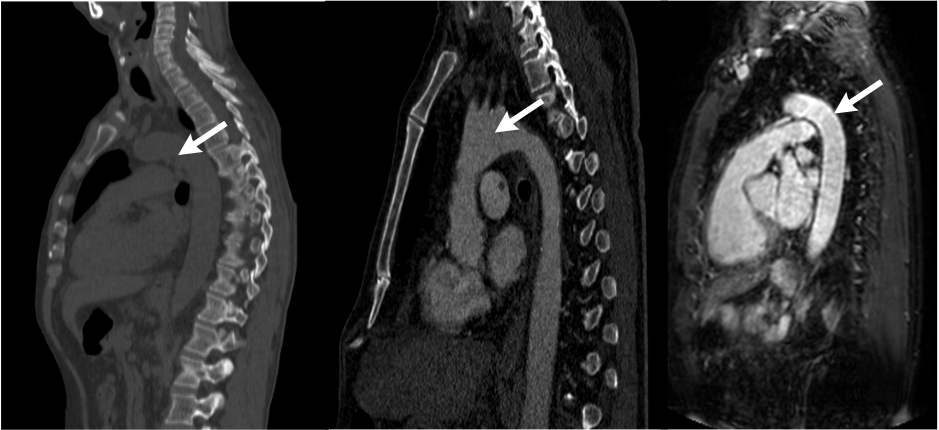}
    \caption{Sagittal views for computed tomography (CT, left), computed tomography angiography (CTA, middle) and magnetic resonance angiography (MRA, right) images. The arrows refer to the aorta, which is most visible (brighter) with a contrast agent (CTA and MRA). Datasets taken from SegTHOR database~(\cite{Trullo_2017, lambert2020segthor}), CAD-PE Challenge [\url{http://www.cad-pe.org/}] and Medical Segmentation Decathlon~(\cite{Simpson_2019}), respectively.}
    \label{fig:CTCTA}
\end{figure}

Medical imaging techniques, such as computed tomography (CT), computed tomography angiography (CTA), magnetic resonance imaging (MRI) and magnetic resonance angiography (MRA), can produce 3D images of the body and are usually employed to examine the patient for aortic diseases, as shown in \autoref{fig:CTCTA}. Depending on the technology, medical images show a range of different internal anatomical structures. The observation of the aorta in the medical images is generally performed by experts, such as vascular radiologists. These need quantitative information about the aortic caliber and, hence, ideally a segmentation of the aorta and its branch vessels from the original medical images (\cite{Erbel_2014}).

\begin{figure}
    \centering
    \includegraphics[width=1.0\linewidth]{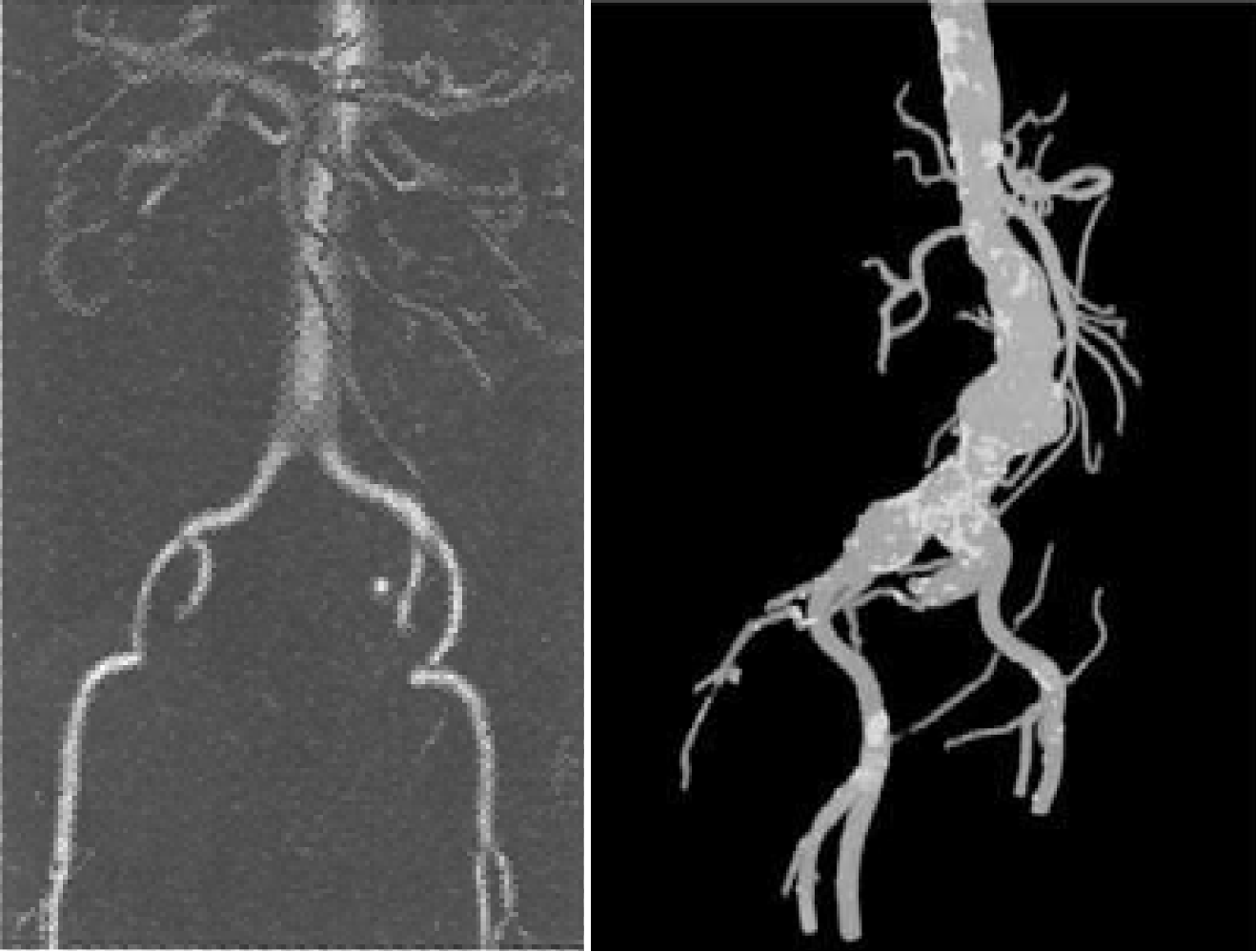}
    \caption{Vesselness filter results for the aortic vessel tree. Left: Result for a contrast enhanced (gadopentetic acid) MRA (magnetic resonance angiography) from \cite{Frangi_1998}. Right: Result for a contrast enhanced CT dataset from \cite{Bauer_2010}.\emph{Per se}, these vesselness filters can be seen as a pre-processing step to a subsequent semantic segmentation that adds labels to the single vessels and vessel sections.}
    \label{fig:Vesselness}
\end{figure}

Manual segmentation of the aorta is a time-consuming task and may lack reproducibility~(\cite{Gamechi_2019,Hahn_2020}). From the last century, researchers have been working on segmenting vessels from raw medical images. In 1998, \cite{Frangi_1998} developed a vessel enhancement filter based on all eigenvalues of the Hessian matrix. In this study the developed filter has been applied to segment the aorta from 3D MRA images (\autoref{fig:Vesselness}, left). In 2000, \cite{Krissian_2000} proposed a model-based approach to reconstruct 3D tubular structures and extract their centerlines. The approach was tested based on real 2D X-ray subtracted angiographies and MRA images of brain vessels. In 2005, \cite{Pock_2005} further developed a robust tube detection filter for 3D centerline extraction. The authors have compared their method with the previous two from \cite{Frangi_1998} and \cite{Krissian_2000}. Finally, in 2010, \cite{Bauer_2010} presented a novel approach for tubular tree structures segmentation, utilizing shape priors
and graph cuts (\cite{Boykov_2001}). A visual comparison of the results from \cite{Frangi_1998} and \cite{Bauer_2010} is demonstrated in \autoref{fig:Vesselness}.

These works can mostly be seen as important pre-processing steps that extract or enhance tubular-like structures in raw (medical) images, which are, in fact, mostly vessels. A particular exception is given by diseases that asymmetrically deform original shape of a blood vessel, like aortic dissections. However, these are rare, and most vessels keep their tubular structure, even when developing a pathology that deform their structure like some aneurysms and stenoses. Nevertheless, \emph{per se} these so-called \emph{vesselness} filters do not provide any semantic information, which means they do not label and identify single vessels or vessel sections.

Further semi-automatic or automatic aorta segmentation methods, however, require experts or ground truth images to evaluate their quality. However, these algorithms can greatly assist clinicians, and thus, they are currently one of the most investigated topics in medical imaging. The goal of this contribution is to review different aorta segmentation algorithms, summarize their strengths and limitations regarding segmentation accuracy, computational complexity and fields of application. 

We give a comprehensive review of state-of-the-art approaches for aorta segmentation. The goal is to help implement better aortic vessel tree segmentation approaches, as well as to: 
\begin{enumerate}
    \item Discuss detailed evaluation criteria for aortic vessel tree segmentation;
    \item Evaluate and conclude the best state-of-the-art approach for this task;
    \item Conclude the main challenges and future directions for this task.

\end{enumerate}
\paragraph{\textbf{Manuscript Outline}}
In the following, \autoref{sec:Aortic Pathologies} provides an overview of different studies on aorta segmentation, with a breakdown of the segmented aortic parts and clinical-grade morbidities. \autoref{tab:aorticPapers} also provides a brief summary of this section. \autoref{sec:algorithms}, instead, provides a review of these papers and classifies them into four main categories based on the algorithms involved: deformable model approaches, tracking approaches, deep learning approaches and additional approaches. Types and amount of input data, evaluation metrics, and other functional parameters are also taken into account, which are presented in detail in \autoref{tab:deformablemodel}, \autoref{tab:tracking}, \autoref{tab:deeplearning}, and \autoref{tab:additional}. Several fields of applications for the segmented aortic volumes are then discussed in \autoref{sec:Clinical Application}, followed by a conclusion with new directions for future work in \autoref{sec:Conclusion and Discussion}.


\paragraph{\textbf{Search Strategy, Inclusion and Exclusion Criteria}}
According to the Systematic Reviews and Meta-analyses Protocols guidelines~\cite{Shamseerg_2015}, we first performed a systematic search in IEEE Xplore, PubMed, ScienceDirect and Google Scholar for the keywords \lq aorta\rq and \lq segmentation\rq. During the search, we retrieved $833$ non-distinct records, and considered six additional papers, which were already known to us. Based on the titles and abstracts, papers with clinical contributions only, such as measurements of the aortic caliber and other medical findings, were excluded (e.g., \cite{Morris_2019}). After this step, we assessed the resulting $131$ distinct papers and excluded $87$ of them. This exclusion was mostly due to the overlap of content (segmentation workflow and approach) with other papers or the lack of quantitative segmentation results. The remaining $44$ papers will be presented within this contribution\footnote{More information on the search strategy and results can be found in the supplementary material.}. To the best of our knowledge, this is the first systematic review that gives a thorough analysis of all published aortic vessel tree segmentation papers.

\section{Aortic Vessel Tree Segmentation: Overview and Challenges}
\label{sec:Aortic vessel tree segmentation: current state}

There are several factors that influence the design of an aortic vessel tree segmentation model, including the type of medical imaging techniques (e.g. CT or MRI), the field-of-view of the radiological examination (e.g., thoracic or abdominal scan), and the underlying segmentation approach. \autoref{tab:aorticPapers} organizes studies of aorta segmentation approaches with respect to the parts of the aortic vessel tree segmented.

\begin{landscape}
    \centering
    \begin{longtable}{c c c c c}
    \caption{Overview of the reviewed publications according to which parts of the aortic vessel tree they segmented (the vessel labels correspond to the ones in~\autoref{fig:AorticVesselTree}) and main diseases involved: Calcifications, aneurysms, or aortic dissections. The publications are ordered and split into four sections according to the other tables in the manuscript: deformable model algorithms (\autoref{tab:deformablemodel}), tracking algorithms (\autoref{tab:tracking}), deep learning algorithms (\autoref{tab:deeplearning}) and additional algorithms (\autoref{tab:additional}).}
    \label{tab:aorticPapers} \\
                            \toprule
    Study& No disease & Calcifications & Aneurysms & Aortic dissections \\  \midrule
    \endfirsthead
    \multicolumn{5}{r}{\autoref{tab:aorticPapers} (continued)}\\
    \toprule
    Study&No disease & Calcifications & Aneurysms & Aortic dissections \\  \midrule
    \endhead
    \bottomrule
    \multicolumn{5}{c}{continued}
    \endfoot
    \bottomrule
    \endlastfoot
    \cite{Rueckert_1997}&1&&&\\
    \cite{Loncaric_2000}&&&4&\\
    \cite{Subasic_2000}&&&4&\\
    \cite{Olabarriaga_2005}&&&4& \\
    \cite{Das_2006}&&&4& \\
    \cite{zhuge_2006}&&&4&\\
    \cite{kim_2007}&&&4&\\
    \cite{Herment_2010}&1,3&&&\\
    \cite{Auer_2010}&&&4,14,15&\\
    \cite{Krissian_2014}&&&&1,2,3,4,5,6,7,12,13,14,15 \\
    \cite{Kurugol2015}&&1,2,3,4&&\\
    \cite{Bustamante_2015}&1,2,3,4,5,6,7&&&\\ 
    \cite{Volonghi_2016}&1,2,3,4,5,6,7&&&\\
    \cite{Gao_2016}&1,2,3,4,14,15&&& \\
    \cite{Duan_2016}&&&&1,2,3,4  \\
    \cite{Morais_2017}&1,2,3,4&&& \\
    \cite{Wang_2017}&&&4,14,15&\\
    \cite{Bidhult_2019}&1&&&     \\  
    \cite{Lareyre_2019}&&&4,11,12,13,14,15&\\
    \cite{Kosasih_2020}&&&4&\\ \midrule
    
    \cite{Subramanyan_2003}&&&4,13,14,15&\\
    \cite{Boskamp_2004}&4,15&&&\\
    \cite{Zheng_2010}&1,2,3,&&&\\
    \cite{Biesdorf_2011}&1,2,3,4&&&\\
    \cite{Mera_2013} &1,2,3,4,5,6,7,10,11,12,13&&& \\ 
    \cite{Xie_2014} &1,2,3,4&&& \\   
    \cite{Selver_2016}&&&4,9,11,12,14,15& \\
    \cite{Tahoces_2019} &1,2,3,4&&&   \\  \midrule
    
    \cite{Trullo_2017}&1,2,3,4&&& \\
    \cite{Linares_2018} &&&4&  \\ 
    \cite{Li_2019} &&&&1,2,3,4  \\  
    \cite{Cao_2019}&&&&1,2,3,4 \\
    \cite{Fantazzini_2020}&1,2,3,4,5,6,7,12,13,14,15&&&   \\  
    \cite{Howard_2020}&1,2,3,4&&&\\
    \cite{Hepp_2020}&1,2,3,4&&&\\
    \cite{Berhane_2020}&1,2,3,4,5,6,7,10&&&\\
    \cite{Hahn_2020}&&&&1,2,3,4\\
    \cite{Chen_2021}&&&&1,2,3,4,5,6,7,10,11,12,13,14,15\\
    \cite{Yu_2021}&&&&1,2,3,4\\
    \cite{Jin_2021}&1,2,3,4,5,6,7&&&\\
    \cite{Zhong_2021}&1,2,3,4,5,6,7&&&\\
    \cite{Cheung_2021}&1&&&\\ \midrule
    
    \cite{de_2002}&&&4&\\
    \cite{Bodur_2007}&&&1,2,3,4&\\
    \cite{Duquette_2012}&& & 4& \\
    \cite{Freiman_2012}&1,2,3,5,6,7&&& \\
    \cite{Bustamante_2015}&1,2,3,5,6,7&&& \\  
    \cite{Gamechi_2019}&1,2,3,4&&& \\
    \bottomrule
 \end{longtable}
\end{landscape}

Based on our review, there are only a few studies focused on semi- or fully automatic segmentation of whole aortic vessel trees~\cite{Shahzad_2015, Chen_2021}. To the best of our knowledge, there could be several reasons behind this.

\begin{itemize}
\item Lack of original datasets~(\cite{Pepe_2020}). Tasks of whole aortic vessel tree segmentation require large amounts of raw data. Scanning should include both thorax and abdomen, which also means a longer scanning duration and higher radiation dose for CTs. Due to the potentially harmful effects of ionizing radiation, this should be limited to retrospective imaging.
\item Lack of segmentation masks as ground truth. The generation of ground truth masks for aortic trees is expensive, as documented by \cite{Hahn_2020}.
\item Image noise as a major problem during automatic segmentation. Newly acquired images might present a different amount of noise compared to the original training sets. Thoracic pain, common in many vasculopathies, often limits a patient's ability to remain still during image acquisition and/or hold their breath, and, together with the movements of the physiological heartbeat can lead to motion artifacts~(\cite{barrett2004artifacts,gallagher2008use}).
\end{itemize}

\section{Aortic Pathologies: Clinical Workflow, Implication and Relevance}
\label{sec:Aortic Pathologies}
As the largest artery in the human body, diseases of the aorta can lead to serious complications~(\cite{Maton_1995}). The aorta and its branches can be involved in several different pathologies and abnormalities: acute aortic syndromes, aortic aneurysms, atherosclerosis, aortic rupture, pseudoaneurysms, inflammatory diseases, genetic diseases as well as congenital abnormalities (\cite{Erbel_2014}). In this section, we introduce some common aortic diseases.

The name of \textbf{acute aortic syndrome (AAS)} refers to aortic pathologies that involve the aortic wall (\cite{tsai2005acute}). These include aortic dissection (AD), intramural hematoma (IMH), and penetrating atherosclerotic ulcer (PAU), the most common of which is AD (85\% to 95\% of all AAS) (\cite{Harris_2012}).

In a classical sense, \textbf{aortic dissection} is the consequence of a tear in the aortic intima. Blood passes through the tear separating the intima from the media and adventitia, creating a false lumen (\cite{Khan_2002}). Evidence showed that the tear often follows a degenerative or necrotic process of the aortic media (\cite{levy2022aortic,osada2018histopathological,wu2013molecular}). Depending on the location of this tear, AD can be classified according to the Stanford classification (\cite{LeMaire_2011}) as a Type A aortic dissection (TAAD) involving any part of the aorta proximal to the origin of the left subclavian artery or a Type B aortic dissection (TBAD), distal to the origin of the left subclavian artery. The diameters and volumes of the aorta, as well as its branches, the presence of mural thrombus, and the shape and area of the tears are crucial for properly characterizing AD (\cite{Rogers_2011}). \cite{Pepe_2020} recently discussed and reviewed the literature on AD.

An \textbf{intramural hematoma} (IMH) is a contained aortic wall hematoma with bleeding in the media without an intimal tear. Acute IMHs appear as focal, crescentic, high-attenuating regions of eccentrically thickened aortic wall on non-contrast CT. IMH have a lower attenuation compared to the aortic lumen on post-contrast CT. IMH will displace atheromatous calcifications into the aortic lumen, while mural thrombus will displace calcifications away from the lumen.  However, occasionally IMH, aortic thrombus and atherosclerotic thickening are difficult to distinguish with CT. In this case, MRI might be a better solution for diagnosis, as the method of dynamic cone gradient-echo sequences can be applied (\cite{Nienaber_2013}).

\textbf{Penetrating atherosclerotic ulcers} are ulcerating atherosclerotic lesions that penetrate the intima into the media of the aortic wall. (\cite{Bolger_2008}). In non-constrast CT images, PAU is similar to IMH. CTA is the preferred technique for the diagnosis of PAU, where the typical finding is a contrast filled out-pouching of the wall of the aorta or into a thickened aortic wall in absence of an intimal flap or false lumen, more commonly located in the descending aorta. The disadvantage of MRI is its limitation in displaying the displacement of the intimal calcifications, which can occurs with PAU.

An \textbf{aortic aneurysm} is a focal or diffuse dilatation of the aorta involving all three layers of the aortic wall. It can be classified into thoracic aortic aneurysms (TAA) and abdominal aortic aneurysms (AAA). TAA can be detected in many locations on the thoracic aorta, and is, in most cases, asymptomatic. TAA is often detected when patients are examined with medical imaging for investigative reasons, which puts higher demands on the capabilities of medical imaging. AAA, as the name suggests, is detected in the abdominal aorta. With the improvement of technology, CT and MRI are now considered better than aortography and the current gold standard for diagnosis, as well as operative assessment of AAA. Diameter, length and curvature of the aorta are especially important for the surgical operation and repair of the aortic aneurysm (\cite{Erbel_2014}). 

\textbf{Aortic atherosclerosis} is caused by endothelium activation and accumulation of lipids in the intimal and medial layers of the aortic wall (\cite{Hager_2007}). The process of atherosclerosis can lead to many diseases, including thrombosis, atherosclerotic aortic occlusion, calcified aorta and coral reef aorta (\cite{Erbel_2014}). Imaging techniques (CT or MRI) are helpful in the diagnosis and treatment of atherosclerotic diseases by providing information about calcifications and plaques.

Furthermore, imaging techniques can be applied in the diagnosis and treatment of other aortic diseases, including traumatic aortic injuries, aortic coarctation, aortitis, connective tissue diseases such as Marfan and Loeys-Dietz syndromes, and more. They are also important for long-term follow-up and monitoring of aortic disease development. Non-invasive examinations help evaluate the conditions of patients and monitor treatment. Advanced segmentation methods can separate the aorta from other organs, and help the clinicians in analyzing the characteristics of patients and reach their diagnosis. 

\section{Computer-aided Aortic Tree Segmentation}
\label{sec:algorithms}
Based on our search, aortic vessel tree segmentation usually focuses on volume segmentation of healthy aortas in CT, CTA or MR images, while some studies also aim to segment specific aortic branches or aortas with different diseases. In this section, we introduced and compared various aorta segmentation algorithms from reviewed studies. 

\subsection{Review Organization}
\label{subsec:metrics}


Various aorta segmentation approaches were identified in this paper. In this subsection, we introduced the taxonymy of approaches and the performance metrics.

The approaches identified in this review can be roughly divided into four classes, namely, deformable models, tracking models, deep learning, and other techniques. We discuss these four classes in the following subsections. In addition, the pre-processing steps, the degree of automation, and the specific segmented parts of aortic vessel tree were also taken into account.

To compare the performance of different approaches from different studies, we mainly focused on their datasets and metrics. For datasets, we paid particular attention to data quality, namely the modalities and the amount of raw images, and the clinical status of patients. During our review, we found that research typically utilizes four methods for collecting the validation data. Most of them obtained the raw medical images from retrospective examinations of hospitals or other specific healthcare facilities, while others also considered publicly available datasets or relied on synthetic images or digital phantoms. Only a few studies prospectively recruited volunteers or patient data for validation.

Different metrics have been proposed for performance evaluation: time taken to complete a segmentation and the accuracy of segmentation results against the standard of practice were most common. Researchers have applied several qualitative and quantitative metrics of segmentation accuracy. In earlier times, the approaches were usually evaluated visually. The researchers invited clinical experts to manually evaluate their segmentation results and/or give a score of segmentation performance. In some studies, researchers might only give examples to show their segmentation results, which makes it difficult to compare the segmentation results with other studies.

Then a metric called volume error (VE) was sometimes applied based on:
\begin{equation}
VE=\frac{|V(A)-V(B)|}{V(A)}
\end{equation}
where $V(A)$ and $V(B)$ are volumes of original and correctly classified images, respectively. This metric can partly represent the segmentation performance, but has problems when a small object is segmented from a relative large background. 

In recent years, more studies have applied another metric called the dice similarity coefficient (DSC)~(\cite{taha2015metrics}), which can be defined as:
\begin{equation}
DSC=\frac{2|A\cap B|}{|A|+|B|}
\end{equation}
where $|A|$ and $|B|$ represent the voxels of aortic vessels in segmentation and ground truth images, respectively. It is essentially a measure of overlap between two sets and can better evaluate the segmentation approaches. 

In addition, the Hausdorff distance (HD), which measures the local maximum distance between two surfaces, is also applied to evaluate the segmentation performance. It can be defined as:

\begin{equation}
HD = \text{max}\{ {\underset{s\in S}{\text{sup}}}{\underset{g\in G}{\text{inf}}} d(s,g), {\underset{g\in G}{\text{sup}}}{\underset{s\in S}{\text{inf}}} d(s,g)\}
\label{eq:hausdorffdistance}
\end{equation}
where $S$ and $G$ are the segmentation results and the ground truth, $d$ represents the Euclidean distance, $sup$ is the supremum and $inf$ is the infimum.

One additional metric called average symmetric distance (ASD) can also help during the evaluation (\cite{Heimann_2009}). It is defined as:
\begin{equation}
\begin{aligned}
ASD = \frac{1}{|P(A)|+|P(B)|}&\left( \sum _{P(A)}d(P_A, P(B)) \right. \\
& \left. +\sum _{P(B)}d(P_B, P(A)) \right) \\
\end{aligned}
\end{equation}
where $P(A)$ and $P(B)$ are the set of points from region A and B; $d(P_A,P(B))$ and $d(P_B,P(A))$ are the shortest distances between any point from region A to region B and vice versa, calculated with the Euclidean distance.

\subsection{Deformable Models}
\label{subsec: Deformable Approaches}

Deformable models (\cite{kass1988snakes}, \cite{Terzopoulos_1991}) apply curves or surfaces to move and fit the contours of images. The goal of the fitting process is usually to minimize an energy function influenced by external and internal forces. External forces pull the curves or surfaces towards the contours, while internal forces are designed to resist deformation and therefore ensure smooth results. Deformable model approaches can further be classified into edge-based and region-based approaches.

As one of the most widely used segmentation algorithms in the early stages, studies based on deformable model algorithms have a long time span. In 1997, \cite{Rueckert_1997} published their research on aorta segmentation based on a deformable model approach. The authors presented a novel algorithm for the tracking of the aorta in cardiovascular MR images. A multiscale medial response function was first applied to roughly locate the aorta. This estimate was then refined with a deformable model, which was defined in a Markov-Random-Field framework. The results have been evaluated with clinical compliance studies to show its ability in clinical applications.

\begin{figure}
    \centering
    \includegraphics[width=0.7\linewidth]{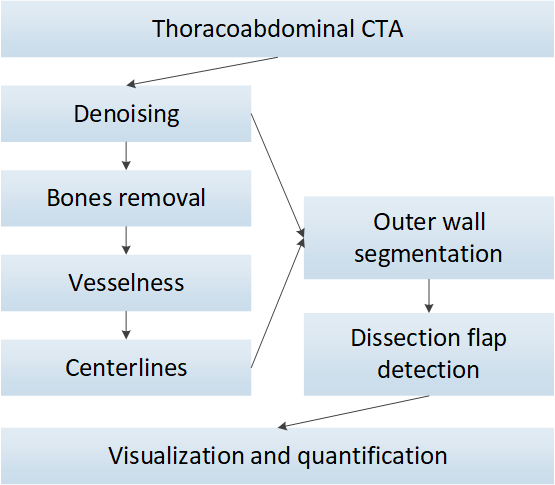}
    \caption{Workflow of aorta segmentation in~\cite{Krissian_2014}. CTA: computed tomography angiography.}
    \label{fig:krissian_2014}
\end{figure}

\cite{Krissian_2014} proposed to use a combination of methods to segment aortas affected by AD. They first applied pre-processing methods to remove image noise and other organs from original images. Then, a multiscale algorithm was applied to extract the possible centerlines. Additional centerlines (false positives) could be present due to smaller vessels or bones missed during the pre-processing step and therefore a manual centerline selection step was always performed. Finally, the selected centerlines were used as initialization points to extract the aortic wall as well as the AD flap with a level-set algorithm. The workflow is shown in \autoref{fig:krissian_2014}. This approach was evaluated with synthetic and real images from three patients with AD. The authors compared the segmentation results of their approach with manually segmented dissections. The comparison showed an average distance of about $0.5$ voxels. However, the lack of data still posed difficulties in making a convincing assessment. Some steps of the complicated pre-processing and manual extraction of centerlines should also be simplified.

\cite{Kurugol2015} proposed a 3D level-set method to segment the aorta in CT images and performed calcification detection. Aorta segmentation was based on a selected region of interest from the original images. The initialization of the aortic surface was based on its approximate tubular shape. At first, a rough surface was created slice by slice within two steps: (1) A circular Hough transform was applied to detect circles slice by slice; (2) the contour of the aorta was selected from all the detected circles based on the posterior probability related to the Hough transform, the spatial smoothness prior and the current slice. Then, a refinement of all the selected aortic circles was performed based on a 3D level-set method, which evolves the shape of the initialized surface to match the real aortic edge. In addition to the segmentation of the aortic wall, they also attempted an automatic detection of calcifications by using a simple threshold-based method to segment plaques with more than 130 Hounsfield units (HU). However, the method has some limitations, including application range and segmentation accuracy. Some of the segmented calcification plaques were actually healthy aortic wall, which is usually located in regions of high HU values. In addition, this method can only be applied to specific CT images with stable HU values, and the threshold value needs to be tuned for different images to get better segmentation results.

Deformable model approaches are widely applied in the field of vessel segmentation~(\cite{Moccia_2018}). One limitation of these algorithms is that deformable models are mainly driven by external forces of image densities. As a result, it is challenging to segment the aorta from original images which have weak intensity gradients or image noise. In addition, the deformable model approaches usually need to be initialized, which means few studies have applied deformable model algorithms to perform a fully automatic segmentation of the aorta. 

A summary of the deformable model algorithms is presented in \autoref{tab:deformablemodel}. From this table, the main advantage of deformable model algorithms is their directness. In the past decades, when insufficient computing resources were available, deformable models could still be applied to segmentation tasks on common hardware. Additionally, this algorithm does not have large requirements for ground truth information, so it's still usable with less data support. However, this type of segmentation algorithm demands stronger contrast within original images. Therefore, all segmentation tasks based on deformable algorithms were achieved after many relatively complex pre-processing steps. Moreover, in these algorithms, the gradient of boundary pixels (voxels) is repeatedly calculated, which means this type of algorithm cannot be directly applied to the original medical images. To resolve it, some studies used prior knowledge of medical imaging to remove other organs and improve image quality (\cite{zhuge_2006},\cite{Lareyre_2019}). Others chose to manually select the regions of interest to improve segmentation efficiency and reduce errors. Despite these problems, the deformanble model algorithms are still widely applied as unsupervised or semi-supervised segmentation methods, especially when they are used mainly for segmentation refinement.

\begin{landscape}

\begin{table*}[ht]\footnotesize
  \centering
    \caption{Summary of deformable model algorithms for aorta segmentation. Note that the number of digits after the decimal points vary, because we used the original numbers provided within the publications.}
    \captionsetup{width=1\linewidth}
  \label{tab:deformablemodel}
  \begin{tabular}{c c c c c c c c c c c c}
                            \toprule
        
    \multirow{2}{*}{Study}&\multicolumn{4}{c}{Method}&\multicolumn{3}{c}{Dataset}&\multicolumn{4}{c}{Evaluation} \\  
    &Algorithm&Pre.&Auto&SVT&Modality&No.&Disease&Duration&RS& Metrics&Statistic \\ \midrule
    \cite{Rueckert_1997}&EM&A&=&=&MRI&87(p)&-&3min&GT&VE&2.21\\
    \cite{Loncaric_2000}&LS&R&-&=&CTA&1(p)&AAA&NA&OB&-&-\\
    \cite{Olabarriaga_2005}&GLM&R&=&=&CTA&17(p)&AAA&41s&GT&VE/MDE&4.5/1.3 \\
    \cite{Das_2006}&AC&D\&M&=&=&CT&7(d)&AAA&NA&GT&DSC&\textgreater85.1\\
    \cite{zhuge_2006}&LS&D&+&=&CTA&20(p)&AAA&7min&GT\&C&VE/MDE&3.5/0.6\\
    \cite{Herment_2010}&DM&R\&A&-&=&PCMR&52(p\&v)&-&17s&GT\&C&DSC&94.5\\
    \cite{Auer_2010}&DM&R&-&=&CTA&11(p)&AAA&NA&C&MDE&-\\
    \cite{Krissian_2014} &AC&D&-&+&CT&3(s)/5(p)&-/AD&8min&OB\&C&MDE&-\\ 
    \cite{Kurugol2015} &LS&R&-&-&CT&45(d)&AC&NA&GT&DSC&92 \\  

    \cite{Volonghi_2016}&LS&D\&R&-&+&PCMR&6(v)&-&30s&GT&MDE/DSC&\textless 1.44/\textgreater88\\
    \cite{Gao_2016}&DM&D&+&+&CTA&36(p)&-&90s&GT\&C&DSC&\textgreater 95\\
    \cite{Duan_2016}&GVF&M&=&+&CT&5(p)&AD&1s&GT&DSC&92.7\\
    \cite{Morais_2017}&BEAS&R&-&=&CT&42(p)&d&73s&GT&DSC/HD&94/1.61\\
    \cite{Wang_2017}&AC&D\&M&=&=&CEMRA&19(p)&AAA&35s&GT\&C&DSC&89.8\\
    \cite{Bidhult_2019}&AC&M\&R&-&=&PCMR&191(p\&v)&-&\textless3.1s&GT&DSC&95.3     \\  
    \cite{Lareyre_2019}&AC&D\&A&+&=&CT&40(p)&AAA&\textless1min&GT&DSC/HD&93/1.78\\
    \cite{Zhao_2022}&MSMR&A&+&=&CTA&35(d)&AD&16.6s&GT\&C&DSC/HD&94.12/2.85\\
    \bottomrule
  \end{tabular}
  \justifying{\\List of abbreviations:\\
Algorithm: EM = energy minimizing, LS = level set, GLM = grey level model, AC = active contour, DM = deformable model, GVF = gradient vector flow, BEAS = B-spline explicit active surfaces, MSMR = morphology-constrained stepwise deep mesh regression.\\ 
Pre-processing before main segmentation process (Pre.): M = manual segmentation of contours, A = (semi) automatic delineation of contours, R = region of interest or start point selection, D = denoising or bone removal, (-) not mentioned.\\ 
Automation according to the whole segmentation process (Auto): (+) fully automatic, (-) semi-automatic, (=) more manual guidance required. Initialization steps that go beyond parameter setting and region of interest or start point selection are also defined as excess manual guidance. \\
Segmented parts of aortic vessel tree (SVT): (+) whole aorta with/without some of its branches, (-) thoracic aorta with/without some of its branches, (=) only part of thoracic aorta. According to \autoref{fig:AorticVesselTree} and \autoref{tab:vesseltreelabels}, whole aorta refers to aortic parts 1-4 and thoracic aorta refers to aortic parts 1-3.\\
Modality: MRI = magnetic resonance imaging, CTA = computed tomography angiography, CT = computed tomography, PCMR = phase-contrast magnetic resonance, CMR = cardiovascular magnetic resonance, CEMRA = contrast enhanced magnetic resonance angiography. \\
Number of data for method validation (No.): p = patients from specific health facilities, d = data from public datasets or challenges, v = volunteers, s = synthetic images.\\
Aortic diseases: (-) no disease, d = different diseases, AAA = abdominal aortic aneurysm, AD = aortic dissection, AC = aortic calcification. \\
Duration: NA = not available / reported, min = minute, s = second.\\
Reference standard (RS): GT = ground truth, OB = human observer, C = comparison with other approaches or literature.\\
Metrics: VE = volume error (\%), DSC = dice similarity coefficient (\%), MDE = mean distance error (mm), HD = Hausdorff distance (mm). }
\end{table*}

\end{landscape}

\subsection{Tracking Models}
\label{subsec: Tracking Approaches}
Tracking approaches usually start from a set of initial seed points and use specific growth algorithms to track the vessels and iteratively select new candidate vessel points. Seed points are either manually selected or automatically defined. Although tracking approaches are widely used in the field of vessel tree segmentation (\cite{Robben_2014,Chen_2016,Biesdorf_2015,Wang_2012,Lee_2017}), most of their applications focus on vessel trees of the brain, retinal vessels or coronary arteries. There are fewer studies which apply tracking algorithms to segment the aortic tree.

\cite{Subramanyan_2003} were among the first to employ tracking approaches for aorta segmentation. The authors proposed a semi-automatic method to track the aortic vessel tree from selected seed points. The fast marching method (\cite{Sethian_2000}) was applied to delineate the blood flow, and the distance transform method was then used to extract centerlines. Finally, the fast marching method was reinitialized in a blood filled region, and the CT volume was subtracted to obtain the contours of the aortic thrombus. This approach was evaluated simply using spatial localization and diameter determination compared with manual and 3D volume rendering methods.

\cite{Biesdorf_2011} proposed a model-based tracking method for aorta segmentation and analysis in dynamic 4D electrocardiogram gated computed tomography angiography (ECG-CTA) images. Firstly, a 3D model of the vessels was created based on the parameters of the vessel width, image blur and voxel position. A coarse aorta segmentation was then performed through a gradual fitting process, which was manually initialized by the vessel model. The segmentation was then refined using a Kalman filter and the vessel model. Finally, the 3D motion of the aorta was determined with intensity-based matching. The authors evaluated their method based on Euclidean distance between manually obtained and method-based centerline positions and vessel diameters. However, these two parameters cannot fully represent the segmentation results, as they assume that the aorta is circular along its certerline.

\begin{figure}
    \centering
    \includegraphics[width=0.7\linewidth]{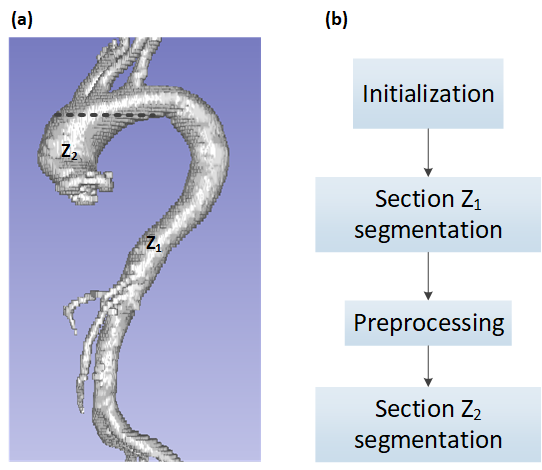}
    \caption{(a) Two parts ($Z_1$ and $Z_2$) of the aorta were segmented respectively with the method of \cite{Mera_2013} (dataset courtesy: Dongyang Hospital). (b) Workflow of the proposed method for aorta segmentation. Section $Z_1$ was first segmented applying a region growing algorithm, then section $Z_2$ was segmented with a level-set algorithm.}
    \label{fig:Mera_2013}
\end{figure}

\cite{Mera_2013} proposed to segment the thoracic aorta in CTA images using a combination of level-set and region growing algorithms. As shown in~\autoref{fig:Mera_2013}, two methods were applied, respectively, to segment the aorta in this study. Parts of the whole aortic tree (partial ascending aorta, aortic arch, descending aorta and some aortic branches) were segmented based on a region growing algorithm. The Hough transform was applied to locate the aorta in the original image and initialize the seed points. Regions were then grown based on the mean and variance of neighboring voxels. The other part of the ascending aorta was then segmented with a level-set algorithm. These two parts were combined to reconstruct the whole aorta. They evaluated their method on ten different CT images and achieved a mean DSC of 0.943 and a Pearson's correlation coefficient (PCC) of 0.972. However, image noise easily affects this method, as the main part of the aorta is segmented by a simple neighbourhood-based region growing algorithm. The combination of segmentation results based on two different methods may also cause incoherence in the connecting surface.

Among current studies on aorta segmentation, the tracking approaches have required the highest continuity in the original images. In this case, the noise in medical images can have a significant impact on the segmentation results. Different approaches, including region growing algorithms, model-based algorithms, ellipse estimation algorithms, and many others, have been applied to track the aortic vessel tree. More algorithms, such as particle filter algorithm~(\cite{Lee_2017}) and statistical Bayesian algorithm~(\cite{Wang_2012}), have been used to track other blood vessels and could potentially be extended to the field of aortic tree segmentation. 

A summary of the current studies on tracking algorithms can be found in~\autoref{tab:tracking}. Similar to deformable model algorithms, tracking algorithms can also be applied as unsupervised or semi-supervised segmentation methods. The core difference is that in tracking algorithms the most attention is paid to image continuity (contours or center lines). In addition, the segmentation process of tracking algorithms is usually faster and easier to visualize. However, these algorithms require even higher image quality than the deformable methods. Thin blood vessels are more likely to be ignored. If the segmentation target is the entire aortic vessel tree, a refinement process should be required after the tracking process.

\begin{landscape}

\begin{table*}[ht]\footnotesize
  \centering
    \caption{Summary of tracking algorithms for aorta segmentation. Note that decimal precision varies, as we report the original numbers provided within the original studies.}
    \captionsetup{width=1\linewidth}
  \label{tab:tracking}
  \begin{tabular}{c c c c c c c c c c c c}
                            \toprule
        
    \multirow{2}{*}{Study}&\multicolumn{4}{c}{Method}&\multicolumn{3}{c}{Dataset}&\multicolumn{4}{c}{Evaluation} \\  
    &Algorithm&Pre.&Auto&SVT&Modality&No.&Disease&Duration&RS& Metrics&Statistic \\ \midrule
    \cite{Subramanyan_2003}&FM&R&-&-&CT&15(p)&AT&\textless90s&C&-&-\\
    \cite{Boskamp_2004}&RG&R&-&=&CTA&1(p)&-&NA&OB&-&-\\
    \cite{Zheng_2010}&PB&-&+&-&CBCT&192(d)&-&1.4s&GT&SPD&1.1\\
    \cite{Biesdorf_2011} & MB&R\&A &-&-&ECG-CTA&30(p)&-&NA&GT&CP&1.53\\
    \cite{Egger_2009}&GB&R\&M/A&=&-&CTA&50(p)&AT&\textless90s&GT&DSC&90.67\\
    \cite{Mera_2013} & RG\&LS&A&+&+&CTA&10(p)&-&NA&GT&PCC&93.83 \\ 
    \cite{Xie_2014} & MB&R&-&-&CT&359(d)&-&NA&GT\&C&DSC/MDE&93.3/1.39 \\   
    \cite{Selver_2016}&3D-PGDF&R&-&=&CTA&19(d)&AAA&80.3s&GT\&C&DSC/MDE&71.07/0.31 \\
    \cite{Tahoces_2019} &ET&-&+&-& CT&380(d)&d&NA&GT&DSC/MDE&95.1/0.9 \\  
    \bottomrule
  \end{tabular}
  \justifying{\\List of abbreviations:\\
Algorithm: FM = fast marching, RG = region growing, PB = part-based tracking, MB = model-based tracking, GB = graph-based tracking, LS = level set, 3D-PGDF = 3D pairwise geodesic distance field, ET = ellipse tracking.\\ 
Pre-processing before main segmentation process (Pre.): M = manual segmentation of contours, A = (semi) automatic delineation of contours, R = region of interest or start point selection, D = denoising or bone removal, (-) not mentioned.\\ 
Automation according to the whole segmentation process (Auto): (+) fully automatic, (-) semi-automatic, (=) more manual guidance required. Initialization steps that go beyond parameter setting and region of interest or start point selection are also defined as excess manual guidance. \\
Segmented parts of aortic vessel tree (SVT): (+) whole aorta with/without some of its branches, (-) thoracic aorta with/without some of its branches, (=) only part of thoracic aorta. According to \autoref{fig:AorticVesselTree} and \autoref{tab:vesseltreelabels}, whole aorta refers to aortic parts 1-4 and thoracic aorta refers to aortic parts 1-3.\\
Modality: CT = computed tomography, CTA = computed tomography angiography, CBCT = cone beam computed tomography, ECG-CTA = electro-cardiogram gated computed tomography angiography.\\
Number of data for method validation (No.): p = patients from specific health facilities, d = data from public datasets or challenges, v = volunteers, s = synthetic images.\\
Aortic diseases: (-) no disease, d = different diseases, AAA = abdominal aortic aneurysm, AT = aortic thrombus. \\
Duration: NA = not available / reported, min = minute, s = second.\\
Reference standard (RS): GT = ground truth, OB = human observer, C = comparison with other approaches or literature.\\
Metrics: SPD = symmetric point-to-mesh distance (mm), CP = mean error of centerline position (mm), DSC = dice similarity coefficient (\%), PCC = Pearson's correlation coefficient (\&), MDE = mean distance error (mm). }
\end{table*}

\end{landscape}

\subsection{Deep Learning}
\label{subsec:Deep Learning}

\cite{Long_2014}, \cite{Ronneberger_2015} and \cite{Cicek_2016} were among the first successful works on medical image segmentation with deep learning. Deep learning approaches have been widely used for the task of medical image segmentation ever since (\cite{Hesamian_2019}). In this survey, deep learning approaches refer to methods that use training data with their respective labeled ground truth to train a convolutional neural network in a supervised manner. 

Deep learning methods are used to extract meaningful features directly from the original data and infer the expected segmentation labels. Thanks to the wide availability of open-source code and libraries, the main advantage of deep learning in medical imaging is its straightforward application for segmentation tasks without tedious manual operations. However, the training of a deep learning model requires a large amount of labelled data, which is often not available in medical imaging. Furthermore, 3D medical images usually come with high resolutions and can often not be directly fed to deep networks due to limited computational resources. As a result, pre-processing steps and patch-based training, e.g. for aortic landmarking (\cite{codari2020deep,schmied2021patch}), are often used to cope with hardware limitations (\cite{Litjens_2017}).

To our knowledge, \cite{Trullo_2017} were the first to apply a deep learning approach for aorta segmentation. The authors used fully convolutional neural networks (FCN) to segment four different human organs, including the aorta, in CT imaging. The authors additionally evaluated two approaches to refine the FCN results: the conditional random fields (CRF) and the SharpMask (SM) architecture~(\cite{Pinheiro_2016}). With a dice score of 0.86, their evaluation showed that the combination of SM and CRF achieved the best accuracy for aorta segmentation.

\begin{figure}
    \centering
    \includegraphics[width=1\linewidth]{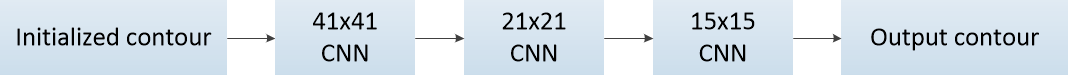}
    \caption{Structure of the proposed cascaded convolutional neural networks (CNNs) for aorta segmentation in~\cite{Li_2019}. Three similar 2D CNN structures with different kernel sizes are applied.}
    \label{fig:Li_2019}
\end{figure}

Two years later, \cite{Li_2019} proposed a cascaded convolutional neural network (CNN) to segment the two aortic lumens in CT scans from patients with AD. The authors first manually extracted 2D cross sections along the aortic centerlines using the Visualization Toolkit (VTK) [https://vtk.org/]. An aortic contour was initialized as an ROI (region of interest) in each cross section. Then, a structure of cascaded 2D CNNs, as shown in~\autoref{fig:Li_2019}, was applied to segment the intima as well as adventitia contours within the ROI. Finally, the segmentation results were reconstructed into 3D volumes using the VTK.

\begin{figure}
    \centering
    \includegraphics[width=0.7\linewidth]{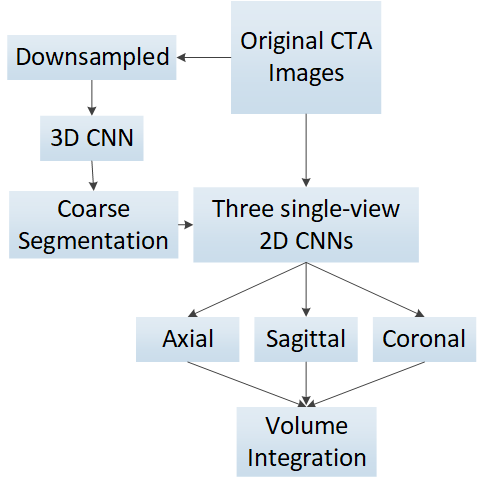}
    \caption{Workflow of the proposed method for aorta segmentation in~\cite{Fantazzini_2020}. A 3D CNN (convolutional neural network) is first used to locate the aorta and three 2D CNNs are then applied to segment the aorta under high resolution based on axial, sagittal and coronal views. An aortic volume segmentation is performed with the integration of these three views. CTA: computed tomography angiography.}
    \label{fig:Fantazzini_2020}
\end{figure}

\cite{Fantazzini_2020} proposed a combination of several CNNs to overcome the problem of high resolution image segmentation. The workflow is shown in~\autoref{fig:Fantazzini_2020}. First, the original images were downsampled, so they could be inputted into a 3D CNN to obtain an aorta segmentation. These results were mainly used as a rough location of the aorta in the original images, and a volume of interest could be cropped. Then, three 2D CNNs were applied to segment the aorta from the cropped volumes in order to perform 2D segmentations based on axial, sagittal and coronal views, respectively. An aortic volume segmentation was performed with the integration of these three views.

Currently there are still relatively few studies that apply deep learning algorithms to segment the aorta. The main reason may be due to the lack of labelled datasets, like ~\cite{Radl2022avtData,radl2022avtPaper}, which has recently been released. The training and evaluation of a deep neural network requires enough training data with ground truth labels. This introduced a strong requirement for manual work of domain experts, like radiologists. In addition, ground truth labels created by different experts may also show a non-negligible inter-operator variability (\cite{Litjens_2017}). Detailed subdivision of different parts of the aorta and its branches, as shown in \autoref{fig:AorticVesselTree}, during the manual annotation, could improve the universality of datasets. 

A summary of the current studies about deep learning algorithms is presented in \autoref{tab:deeplearning}. These deep learning-based approaches have only become popular in recent years, but have shown high potential~(\cite{shen2017deep}). Compared with the previous two methods, deep learning-based approaches are often fully automatic. End users can simply provide the original medical images as input and obtain a segmentation without manual initialization or selection of seed points. A well trained deep learning model can directly output the segmentation results. In addition, it can be applied as a pixel-wise segmentation method to segment thin blood vessels and be utilized with medical images with a lot of noise. However, as a supervised learning algorithm, a well-trained deep learning model typically requires a large set of original images with ground-truth masks. It makes data collection and processing the main challenge of deep learning approaches~(\cite{shen2017deep,Pepe_2020}). 

\begin{landscape}

\begin{table*}[ht]\footnotesize
  \centering
    \caption{Summary of deep learning algorithms for aorta segmentation. Note that the number of digits after the decimal points vary, because we used the original numbers provided within the publications.}
    \captionsetup{width=1\linewidth}
  \label{tab:deeplearning}
  \begin{tabular}{c c c c c c c c c c c c}
                            \toprule
        
    \multirow{2}{*}{Study}&\multicolumn{4}{c}{Method}&\multicolumn{3}{c}{Dataset}&\multicolumn{4}{c}{Evaluation} \\  
    &Algorithm&Pre.&Auto&SVT&Modality&No.&Disease&Duration&RS& Metrics&Statistic \\ \midrule
    \cite{Trullo_2017}&SM\&CRF&N&+&-&CT&30(p)&-&NA&GT&DSC&86\\
    \cite{Linares_2018} & HED&-&+&=&CTA&13(p) & AT &NA &GT\&C&DSC&82  \\ 
    \cite{Cao_2019}&3D U-Net&N\&Aug&+&-&CTA&276(p)&AD&31.1s&GT&DSC&93 \\
    \cite{Fantazzini_2020} &U-Net &N\&R\&Aug&+&+&CTA&80(p)&AAA&25s&GT&DSC&93\\  
    \cite{Howard_2020}&HRNet&N&+&-&CMR&575(p)&-&NA&GT&DSC&92.9\\
    \cite{Berhane_2020}&3D U-Net&N\&D&+&-&4D flow MRI&1193(p)&-&0.44s&GT&DSC/HD&95.1/2.8\\
    \cite{Hahn_2020}&TernausNet&-&+&-&CT&153(p)&AD&\textless 4min&GT&DSC&94.9\\
    \cite{Chen_2021}&DenseNet&N&-&+&CT&120(p)&AD&14.2s&GT&DSC&97\\
    \cite{Yu_2021}&3D U-Net&R\&D&-&+&CTA&139(p)&AD&21.7min&GT&DSC&95.8\\
    \cite{Jin_2021}&V-Net \& PF&Aug&+&-&CT \& CTA&80(d)&-&NA&GT\&OB&VE&\textless 0.255\\
    \cite{Zhong_2021}&AG U-Net&N\&P&+&-&CTA&194(p)&-&NA&GT&DSC&96.6\\
    \cite{Sieren_2022}&3D U-Net&-&+&=&CTA&191(p)&d&60.5s&GT&DSC&95\\
    \bottomrule
  \end{tabular}
  \justifying{\\List of abbreviations:\\
Algorithm: SM = Sharp Mask, CRF = Conditional Random Fields, HED = Holistically-Nested Edge Detection, HRNet =  High Resolution Net, PF = Particle Filter, AG U-Net = Attention-Gated U-Net.\\ 
Pre-processing before main segmentation process (Pre.): R = region of interest or start point selection, Aug = data augmentation, N = normalization of data (value and size), A = (semi) automatic delineation of contours, R = region of interest or start point selection, P = patches division, D = denoising or bone removal, (-) not mentioned.\\ 
Automation according to the whole segmentation process (Auto): (+) fully automatic, (-) semi-automatic, (=) more manual guidance required. Initialization steps that go beyond parameter setting and region of interest or start point selection are also defined as excess manual guidance. \\
Segmented parts of aortic vessel tree (SVT): (+) whole aorta with/without some of its branches, (-) thoracic aorta with/without some of its branches, (=) only part of thoracic aorta. According to \autoref{fig:AorticVesselTree} and \autoref{tab:vesseltreelabels}, whole aorta refers to aortic parts 1-4 and thoracic aorta refers to aortic parts 1-3.\\
Modality: CT = computed tomography, CTA = computed tomography angiography, CMR = Cardiovascular magnetic resonance, CBCT = cone beam computed tomography, ECG-CTA = electro-cardiogram gated computed tomography angiography.\\
Number of data for method validation (No.): p = patients from specific health facilities, d = data from public datasets or challenges, v = volunteers, s = synthetic images.\\
Aortic diseases: (-) no disease, d = different diseases, AT = aortic thrombus, AD = aortic dissection, AAA = abdominal aortic aneurysm. \\
Duration: NA = not available / reported, min = minute, s = second.\\
Reference standard (RS): GT = ground truth, OB = human observer, C = comparison with other approaches or literature.\\
Metrics: DSC = dice similarity coefficient (\%), HD = Hausdorff distance (mm), VE = volume error(\%). }
\end{table*}

\end{landscape}

\subsection{Other Techniques}
\label{subsec: Additional Approaches}

In addition to the three main categories, some studies applied other algorithms for aorta segmentation tasks, although less frequently. Due to the relatively small number of studies based on similar algorithms, these approaches are introduced in this additional subsection.

\cite{de_2002} presented an automatic method based on active shape models (ASM). They applied this method to segment thrombus regions in CTA images of abdominal aortic aneurysms. The original ASM approach was improved with a $k$ nearest neighbour ($k$NN) method, which was used to obtain the probability density estimation that any given profile belongs to the aortic boundary. The comparison of segmentation results of both the original and the improved ASM method showed that the introduction of $k$NN significantly reduced the leave-one-out error from 2.2 to 1.6 mm.

\begin{figure}
\centering
\includegraphics[width=0.7\linewidth]{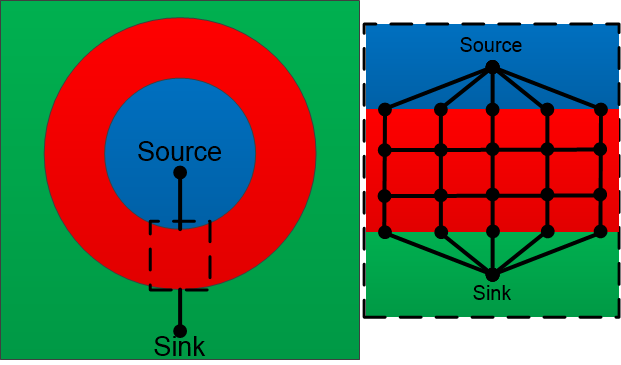}
    \caption{Schematic representation of an initialized aortic area (\cite{Duquette_2012}). Voxels inside the blue area are all connected to the source, and those in green are connected to the sink. Voxels in the middle (red) are connected to their six neighbouring voxels.}
    \label{fig:Duquette_2012}
\end{figure}

\cite{Duquette_2012} applied a graph cut algorithm to segment the aortic volume~(\cite{Boykov_2001}) in 3D CT and MR images. This method can also be seen as an assistance to manual segmentation: The users roughly select the aortic contour in several slices of the 3D images. As shown by~\autoref{fig:Duquette_2012}, the original volume would be divided into three sections: inner section, neutral section, and outer section. Voxels in the inner section (blue) are all connected to the source, while voxels of outer section (green) are all connected to the sink. The graph cut algorithm deals with the voxels in the neutral section (red) to perform an accurate segmentation. An edge capacity of two neighbouring voxels $a$ and $b$ is calculated with:
\begin{equation}
Capacity = exp\left(-\frac{f_{ab}|I_a-I_b|^{\gamma}}{2\sigma ^2}\right),
\end{equation}
where $I$ is the voxel intensity, $\gamma$ is a parameter for gradient-based gamma correction, and $f$ is a binary function, which prevents the segmentation of edges of neighbouring organs. This equation takes into account the gradient information to suit different voxel densities/intensities of CT and MR images. In areas with stronger gradients, the connections between voxels were cut to generate the contours. A manual refinement step usually followed the graph cut method. The authors compared their method against manual segmentations performed by four domain experts. Their comparison showed that the results are sufficiently comparable to the ground truth, which led them to the conclusion that the method can be applied to computer-assisted manual segmentation. An automatic or semi-automatic application of this algorithm can also be realized.

In \cite{Gamechi_2019}, a combination of imaging techniques was applied to segment the aorta in CT images. At first, pre-processing was used to remove the heart, lung and bones from the original CT images. Then, the authors used a multi-atlas-based method~(\cite{Kirili_2010}) to coarsely locate the aorta. The preprocessed images were registered to the target model, and ten of them were selected for the following approaches. Aortic centerlines were traced based on the minimum cost path algorithm. Finally, a graph cut segmentation algorithm was applied to accurately segment the aortic surface based on the traced centerlines. The segmentation results were evaluated based on 100 CT images with manually segmented ground truths. This approach has achieved an average DSC of $0.95\pm 0.01$ and a MSD of $0.56\pm 0.08$ mm.

There are several algorithms applied to segment vessels, but not yet specifically used in semantic aorta segmentation. \cite{Li_2006} developed an optimal surface detection method to detect multiple interacting surfaces, which was applied to segmentation tasks on more than 300 computer-generated 3D images. \cite{Egger_2007} proposed a fast vessel centerline extraction method, which can also be applied for aorta segmentation tasks. \cite{Bruyninckx_2010} applied a global optimization algorithm to segment liver portal veins in CT images. \cite{Sreejini_2015} applied a particle swarm optimization algorithm to segment retinal vessels. \cite{Tang_2012} used a minimum cost path algorithm to segment carotids in MRI. These segmentation algorithms could potentially be extended to the field of aortic vessel tree segmentation.

A summary of additional algorithms is presented in \autoref{tab:additional}.

\begin{landscape}

\begin{table*}[ht]\footnotesize
  \centering
    \caption{Summary of additional algorithms for aorta segmentation. Note that the number of digits after the decimal points vary, because we used the original numbers provided within the publications.}
    \captionsetup{width=1\linewidth}
  \label{tab:additional}
  \begin{tabular}{c c c c c c c c c c c c}
                            \toprule
        
    \multirow{2}{*}{Study}&\multicolumn{4}{c}{Method}&\multicolumn{3}{c}{Dataset}&\multicolumn{4}{c}{Evaluation} \\  
    &Algorithm&Pre.&Auto&SVT&Modality&No.&Disease&Duration&RS& Metrics&Statistic \\ \midrule
    \cite{deBruijne_2003}&ASM&A&-&=&CTA&23(p)&AAA&25s&C&VE&5.9\\
    \cite{Bodur_2007}&IA &R&-&-&CTA&4(P)&AAA&10min&OB&-&-\\
    \cite{Duquette_2012}&GC&R&-&=&CT\& MRI&44(p)&AAA&\textless 1min&GT&HD&\textless3.53 \\
    \cite{Freiman_2012} &WS&D&-&=&CTA&56(d)\&15(p)&-&122s&GT\&C&DSC&84.5 \\
    \cite{Bustamante_2015} &AM&-&+&+&CMR&11(v)/10(p)&-/d&NA&OB&-&- \\  
    \cite{Gamechi_2019}&MR&-&+&-&CT&16(p)&-&NA&GT&DSC&95\\
    
    \bottomrule
  \end{tabular}
  \justifying{\\List of abbreviations:\\
Algorithm: ASM = Active Shape Models, IA = Isoperimetric algorithm, GC = Graph Cut theory, WS = Watershed-based segmentatoin, AM = atlas-based model, MR = Multi-atlas registration.\\ 
Pre-processing before main segmentation process (Pre.): R = region of interest or start point selection, Aug = data augmentation, N = normalization of data (value and size), A = (semi) automatic delineation of contours, R = region of interest or start point selection, P = patches division, D = denoising or bone removal, (-) not mentioned.\\ 
Automation according to the whole segmentation process (Auto): (+) fully automatic, (-) semi-automatic, (=) more manual guidance required. Initialization steps that go beyond parameter setting and region of interest or start point selection are also defined as excess manual guidance. \\
Segmented parts of aortic vessel tree (SVT): (+) whole aorta with/without some of its branches, (-) thoracic aorta with/without some of its branches, (=) only part of thoracic aorta. According to \autoref{fig:AorticVesselTree} and \autoref{tab:vesseltreelabels}, whole aorta refers to aortic parts 1-4 and thoracic aorta refers to aortic parts 1-3.\\
Modality: CT = computed tomography, CTA = computed tomography angiography, CMR = Cardiovascular magnetic resonance, CBCT = cone beam computed tomography, ECG-CTA = electrocardiogram gated computed tomography angiography.\\
Number of data for method validation (No.): p = patients from specific health facilities, d = data from public datasets or challenges, v = volunteers, s = synthetic images.\\
Aortic diseases: (-) no disease, d = different diseases, AT = aortic thrombus, AD = aortic dissection, AAA = abdominal aortic aneurysm. \\
Duration: NA = not available / reported, min = minute, s = second.\\
Reference standard (RS): GT = ground truth, OB = human observer, C = comparison with other approaches or literature.\\
Metrics: DSC = dice similarity coefficient (\%), HD = Hausdorff distance (mm), VE = volume error(\%). }
\end{table*}

\end{landscape}

\section{Clinical Application}
\label{sec:Clinical Application}
In \autoref{sec:algorithms}, we described the main algorithms applied to segment the aorta in medical images. In this section, we introduce some studies on the clinical applications of these approaches, which usually include aortic disease detection (AD, calcification, etc.) and measurement of aortic parameters (diameter, volume, etc.).

\begin{figure}
    \centering
    \includegraphics[width=1.0\linewidth]{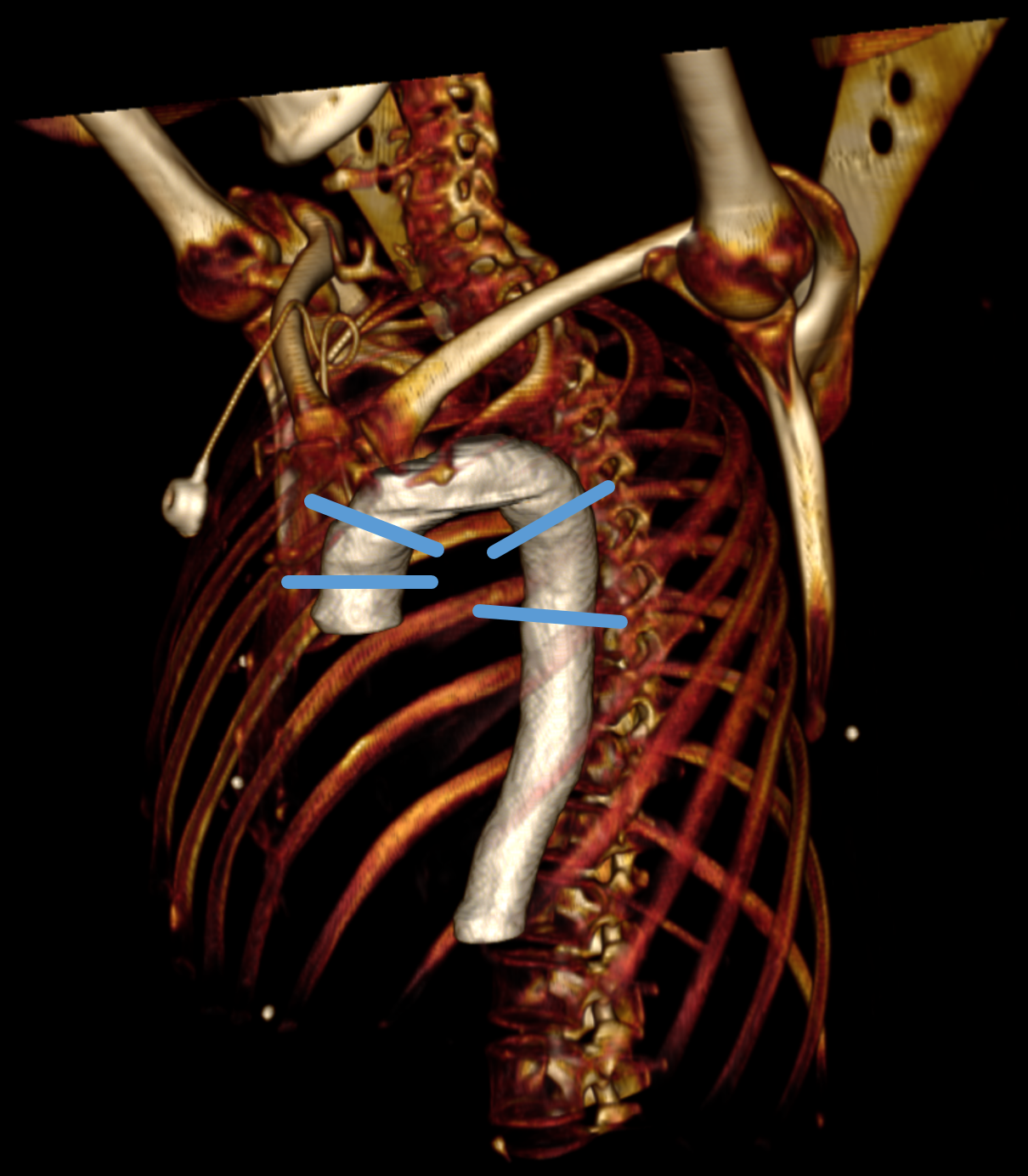}
    \caption{Aorta segmentation results (SegTHOR database~(\cite{Trullo_2017})), were used by~\cite{Gamechi_2019} to measure aortic diameters at different aortic points (blue lines, image rendered with InVesalius~(\cite{amorim2015invesalius}).}
    \label{fig:Gamechi_2019}
\end{figure}

An approach for aorta segmentation in CT images, proposed by~\cite{Gamechi_2019}, was introduced in~\autoref{subsec: Additional Approaches}. The authors introduced a different application of the proposed segmentation method. As shown in~\autoref{fig:Gamechi_2019}, aortic diameter measurements were extracted from multiple cross-sectional slices from segmentation results. Evaluation has shown a small error between automatically and manually obtained aortic diameters, which means this automatic approach can support the clinical analysis of aortas based on original CT images.\footnote{\cite{Egger_2007_preoperative, Egger_2012} also extracted aortic diameter measurements for minimally invasive stent planning and \cite{Jianning} extracted aortic diameters in aortic dissections, like the true and false lumens, and the length of the proximal landing zone for treatment planning.}

\cite{Krissian_2014} proposed a method to perform aorta segmentation in CTA images of ADs. This method was developed to segment the dissection flap and support clinicians with the detection and analysis of ADs.

\begin{figure}
    \centering
    \includegraphics[width=0.7\linewidth]{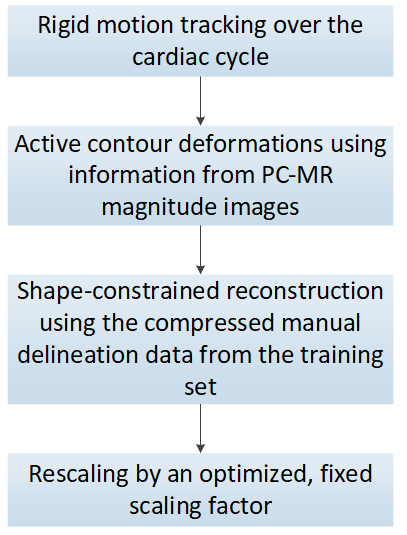}
    \caption{Workflow of the automatic modification of the method in~\cite{Bidhult_2019}. PC-MR: phase-contrast magnetic resonance.}
    \label{fig:Bidhult_2019}
\end{figure}

\cite{Bidhult_2019} proposed an approach based on the active contour algorithm to segment vessel volumes from phase-contrast magnetic resonance images. The method was initialized on manually selected vessels of interest, followed by steps of automatic modification, as shown in~\autoref{fig:Bidhult_2019}. This method was evaluated with 151 human aortas and showed an average DSC of 0.953. In the same study, the algorithm was also applied to measure the blood flow volume and quantify the shunt volume, which demonstrated its clinical applicability.

\cite{Auer_2010} proposed a method based on snake models to segment a part of the human aorta, followed by additional steps for finite element mesh generation. \cite{Erhart_2014} applied this method to perform finite element analysis (FEA) on different kinds of abdominal aortic aneurysms. The geometrical parameters of aortic diameters and volumes were obtained from the segmented aorta, and the mechanical parameters of Peak Wall Stress (PWS), Peak Wall Rupture Index (PWRI) and Rupture Risk Equivalent Diameter (RRED) were calculated based on the homogeneous model properties of wall thickness, mesh size and connective tissue characterization. FEA was performed on different datasets of asymptomatic, symptomatic and ruptured AAAs and the results were compared to support the clinical detection and analysis of AAA.

\section{Conclusion and Discussion}
\label{sec:Conclusion and Discussion}
In this review, we presented a summary of the literature for aortic vessel tree segmentation. We introduced the main aortic pathologies, as well as their clinical treatments, to show the importance of medical imaging techniques for the detection and analysis of aortic diseases. Then, we listed studies of aortic vessel tree segmentation and classified them into four categories: deformable model approaches, tracking approaches, deep learning approaches and additional approaches. The advantages and disadvantages of these approaches were analyzed. In addition, we introduced algorithms that were applied for segmentation tasks of other human vessels, but were not yet used to segment the aorta. Possible applications of segmented aortic volumes were described to give these segmentation approaches more clinical significance.

\subsection{Current State}

Based on our analysis, the distribution of different algorithm categories has changed over time. In the last three years, more studies have applied deep learning approaches to aorta segmentation tasks, while before, more studies applied deformable model and tracking approaches. 

Deformable model and tracking algorithms usually involve multiple steps to automatically or semi-automatically segment aortas or aortic branches from original medical images. These approaches have no requirements for large amounts of data, although annotated datasets are still needed for the evaluation. Each study may segment different parts of aorta, which poses a problem when evaluating and comparing different approaches. 

With the development of deep learning algorithms, as well as the increasing number of datasets relating to the aorta, deep learning approaches have played a more significant role in the field of aorta segmentation tasks in recent years. The main advantages of deep learning approaches are the fully-automatic process and its pixel-wise segmentation performance, which makes it very straightforward to obtain segmentation results. However, a well-trained deep learning model needs a large amount of training data, which is relatively hard to obtain in medical imaging. To our best knowledge, the datasets that currently exist for aorta segmentation, usually only include the main part of the aorta (ascending aorta, aortic arch and part of the abdominal aorta), while other parts of the aorta and aortic branches are ignored. In addition, the training of deep learning models is a time-consuming process, which costs a lot of computing resources. Finally, \autoref{tab:aorticPapers} provides a compact overview of the reviewed papers according to which parts of the aortic vessel tree they segmented, and if and what kind of diseases, like calcification, aneurysm or dissection, was involved in these works. It also reveals that no work has targeted the intercostal arteries (label 8 in \autoref{fig:AorticVesselTree}) so far, which makes sense, because in most scans they are too small to be clearly defined, even manually, according to our medical partner. Nevertheless, we included them in our aortic vessel tree to be consistent with the original publication ~(\cite{Du_2015}). In addition, future scanner generations may visualize them more reliably, and recent works already address this issue, for example, for intra-operative guidance during thoracic endovascular repair~(\cite{koutouzi20173d}).

During our review, some issues emerged when comparing these aorta segmentation approaches from different studies. Although all studies introduced their validation datasets, most were based on private data from hospitals or medical institutes. No information is available about whether the different original images had similar qualities, and if their ground truth segmentation was based on similar standards, and what was the interoperator variability between ground truth annotators. This created difficulties in comparing the performance of their approaches. Additionally, as mentioned earlier, their segmentation results were evaluated using various metrics, which are all statistical tools to measure the segmentation performance, but have different emphases and units. Therefore, it is impossible to objectively compare segmentation accuracy between studies that applied two different assessment metrics based only on published information.

In terms of methods, earlier research focused only on parts of the aortic vessel tree, and a certain degree of human interaction was necessary. Recently, more studies have segmented the aorta and its major branches, and also more studies have applied prior knowledge or deep learning algorithms to develop fully automated segmentation approaches.

Regarding the dataset, a variety of medical images have been collected for segmentation tasks in recent studies compared to previous ones. In addition, the amount of data collected for research, especially in deep learning studies, has also increased significantly.

Assessment methods have also changed. While it was initially difficult to compare segmentation approaches between different studies, in recent years, some major metrics (DSC, HD, MDE) have been widely applied. Researchers have consciously compared their methods with others, which accelerates the development of segmentation algorithms. Besides, some studies still have not mentioned their segmentation duration, which is actually an important factor considering the application of their segmentation methods.

Based on our review, currently the main challenges for aortic vessel tree segmentation are:
\begin{itemize}
\item Lack of common original datasets, which leads to difficulties in comparing the segmentation performance of different approaches;
\item There is no gold standard for aorta segmentation tasks, which also causes difficulties when evaluating and comparing different segmentation approaches;
\item Current annotation tools for medical images are not efficient enough to massively generate ground truth segmentation for training and evaluation;
\item Medical images with high resolution showing the whole aorta cannot, in general, be directly inputted into neural networks. Solutions for this problem, like downsampling, usually reduce the accuracy of the segmentation results.
\end{itemize}

Finally, there is no common definition for a \emph{stopping criteria}, especially for the branching vessels. This can differ between scans due to the scan quality, in-plane resolution, slice thickness, field of view (FOV), contrast agent (distribution), etc. For an quantitative evaluation, in particular for common segmentation metrics, however, a \emph{stopping criteria} should be in a similar range for all scans to enable valid error calculating over the same length of each branching aortic artery.

\subsection{Future Directions}
Our study suggests that deep learning approaches have advantages in aorta segmentation tasks over traditional methods, especially considering automation and generalisation. The main limitation of these algorithms is the lack of datasets and corresponding ground truth annotations, which, when made public to the research community, would also enable more groups to work on this problem and host public biomedical challenges, like the SegTHOR Challenge~(\cite{DBLP:conf/isbi/2019segthor}), which included the main aorta in the thorax, but extended to the whole aortic vessel tree (including the abdominal aorta and the iliac arteries).

To the best of our knowledge, no approach has provided a concise and accurate solution for the segmentation of aortic vessel trees. Some future directions for these tasks could be:

\begin{itemize}
    \item More datasets with well-annotated ground truths should be established and shared to help evaluate and compare different kinds of segmentation approaches;
    \item Classical approaches can be combined with deep learning algorithms to simplify the segmentation process and improve the automation;
    \item To avoid quality loss due to downsampling full body scans of the aorta to input them into neural networks, patch-based approaches should be investigated for the aorta;
    \item More public challenges, e.g. on Grand Challenge (https://grand-challenge.org/), that target aortic vessel tree segmentation tasks should be held to better compare and evaluate existing approaches. 

\end{itemize}
\subsection{Final Remarks}
The aorta is the main artery of the human body, and its diseases pose a significant threat to human life. Computer-aided aorta segmentation is an important step in the detection and analysis of aortic diseases. We conducted a comprehensive introduction of existing aorta segmentation algorithms, as well as some vessel segmentation approaches. Furthermore, we identified the current challenges and future directions in this field. We believe our review provides a summarized reference for researchers who work or want to work in aorta-related medical imaging. In this regards, \autoref{tab:aorticPapers} provides a compact overview for researchers, which parts (but also diseases) of the aortic vessel tree have already been targeted by contributions, but also which have not yet been targeted by the research community. This enables researchers to quickly identify under-researched areas and applications within the aortic vessel tree.

The contribution of our review is threefold. We
\begin{itemize}
    \item provided an overview of aortic diseases where aorta segmentation approaches can play a significant role;

    \item conducted a comprehensive introduction and comparison of existing aorta segmentation algorithms;

    \item provided directions for future work, including the development of segmentation technologies and the expansion of its application domain.
\end{itemize}

\section*{Acknowledgments}
%
This work received funding from the TU Graz Lead Project (Mechanics, Modeling and Simulation of Aortic Dissection), the Austrian Science Fund (FWF) KLI 678-B31: \lq enFaced: Virtual and Augmented Reality Training and Navigation Module for 3D-Printed Facial Defect Reconstructions', and was supported by CAMed (COMET K-Project 871132), which is funded by the Austrian Federal Ministry of Transport, Innovation and Technology (BMVIT), the Austrian Federal Ministry for Digital and Economic Affairs (BMDW) and the Styrian Business Promotion Agency (SFG). AFF was supported by the Royal Academy of Engineering (INSILEX CiET1819\textbackslash 19), and Engineering and Physical Sciences Research Council (TUSCA EP/V04799X/1). Further, we acknowledge CHB (Centre Henri Becquerel, 1 rue d'Amiens, 76038 Rouen, France). Finally, a collection of CTA scans including aortic vessel trees with corresponding segmentations (\cite{Radl2022avtData, radl2022avtPaper}), can be found and freely accessed on Figshare for research or informational purposes: \\
\href{https://figshare.com/articles/dataset/Aortic_Vessel_Tree_AVT_CTA_Datasets_and_Segmentations/14806362}{Aortic Vessel Tree (AVT) CTA Datasets and Segmentations}

%

\bibliographystyle{model2-names.bst}
\biboptions{authoryear}
\bibliography{refs}

\begin{thebibliography}{118}
\expandafter\ifx\csname natexlab\endcsname\relax\def\natexlab#1{#1}\fi
\providecommand{\url}[1]{\texttt{#1}}
\providecommand{\href}[2]{#2}
\providecommand{\path}[1]{#1}
\providecommand{\DOIprefix}{doi:}
\providecommand{\ArXivprefix}{arXiv:}
\providecommand{\URLprefix}{URL: }
\providecommand{\Pubmedprefix}{pmid:}
\providecommand{\doi}[1]{\href{http://dx.doi.org/#1}{\path{#1}}}
\providecommand{\Pubmed}[1]{\href{pmid:#1}{\path{#1}}}
\providecommand{\bibinfo}[2]{#2}
\ifx\xfnm\relax \def\xfnm[#1]{\unskip,\space#1}\fi
\bibitem[{Amorim et~al.(2015)Amorim, Moraes, Silva and
  Pedrini}]{amorim2015invesalius}
\bibinfo{author}{Amorim, P.}, \bibinfo{author}{Moraes, T.},
  \bibinfo{author}{Silva, J.}, \bibinfo{author}{Pedrini, H.},
  \bibinfo{year}{2015}.
\newblock \bibinfo{title}{Invesalius: An interactive rendering framework for
  health care support}, in: \bibinfo{booktitle}{International symposium on
  visual computing}, \bibinfo{organization}{Springer}. pp.
  \bibinfo{pages}{45--54}.
\bibitem[{Auer and Gasser(2010)}]{Auer_2010}
\bibinfo{author}{Auer, M.}, \bibinfo{author}{Gasser, T.C.},
  \bibinfo{year}{2010}.
\newblock \bibinfo{title}{Reconstruction and finite element mesh generation of
  abdominal aortic aneurysms from computerized tomography angiography data with
  minimal user interactions}.
\newblock \bibinfo{journal}{IEEE Transactions on Medical Imaging}
  \bibinfo{volume}{29}, \bibinfo{pages}{1022--1028}.
\newblock \DOIprefix\doi{10.1109/TMI.2009.2039579}.
\bibitem[{Barrett and Keat(2004)}]{barrett2004artifacts}
\bibinfo{author}{Barrett, J.F.}, \bibinfo{author}{Keat, N.},
  \bibinfo{year}{2004}.
\newblock \bibinfo{title}{Artifacts in ct: recognition and avoidance}.
\newblock \bibinfo{journal}{Radiographics} \bibinfo{volume}{24},
  \bibinfo{pages}{1679--1691}.
\bibitem[{Bauer et~al.(2010)Bauer, Pock, Sorantin, Bischof and
  Beichel}]{Bauer_2010}
\bibinfo{author}{Bauer, C.}, \bibinfo{author}{Pock, T.},
  \bibinfo{author}{Sorantin, E.}, \bibinfo{author}{Bischof, H.},
  \bibinfo{author}{Beichel, R.}, \bibinfo{year}{2010}.
\newblock \bibinfo{title}{Segmentation of interwoven 3d tubular tree structures
  utilizing shape priors and graph cuts}.
\newblock \bibinfo{journal}{Medical image analysis} \bibinfo{volume}{14},
  \bibinfo{pages}{172--184}.
\bibitem[{Berhane et~al.(2020)Berhane, Scott, Elbaz, Jarvis, McCarthy, Carr,
  Malaisrie, Avery, Barker, Robinson, Rigsby and Markl}]{Berhane_2020}
\bibinfo{author}{Berhane, H.}, \bibinfo{author}{Scott, M.},
  \bibinfo{author}{Elbaz, M.}, \bibinfo{author}{Jarvis, K.},
  \bibinfo{author}{McCarthy, P.}, \bibinfo{author}{Carr, J.},
  \bibinfo{author}{Malaisrie, C.}, \bibinfo{author}{Avery, R.},
  \bibinfo{author}{Barker, A.J.}, \bibinfo{author}{Robinson, J.D.},
  \bibinfo{author}{Rigsby, C.K.}, \bibinfo{author}{Markl, M.},
  \bibinfo{year}{2020}.
\newblock \bibinfo{title}{Fully automated 3d aortic segmentation of 4d flow mri
  for hemodynamic analysis using deep learning}.
\newblock \bibinfo{journal}{Magn Reson Med.} \DOIprefix\doi{10.1002/mrm.28257}.
\bibitem[{Bidhult et~al.(2019)Bidhult, Hedström, Carlsson, Töger,
  Steding-Ehrenborg, Arheden, Aletras and Heiberg}]{Bidhult_2019}
\bibinfo{author}{Bidhult, S.}, \bibinfo{author}{Hedström, E.},
  \bibinfo{author}{Carlsson, M.}, \bibinfo{author}{Töger, J.},
  \bibinfo{author}{Steding-Ehrenborg, K.}, \bibinfo{author}{Arheden, H.},
  \bibinfo{author}{Aletras, A.H.}, \bibinfo{author}{Heiberg, E.},
  \bibinfo{year}{2019}.
\newblock \bibinfo{title}{A new vessel segmentation algorithm for robust blood
  flow quantification from two-dimensional phase-contrast magnetic resonance
  images}.
\newblock \bibinfo{journal}{Clinical Physiology and Functional Imaging}
  \bibinfo{volume}{39}, \bibinfo{pages}{327--338}.
\newblock \DOIprefix\doi{10.1111/cpf.12582}.
\bibitem[{Biesdorf et~al.(2015)Biesdorf, Worz, Tengg-Kobligk, Rohr and
  Schnorr}]{Biesdorf_2015}
\bibinfo{author}{Biesdorf, A.}, \bibinfo{author}{Worz, S.},
  \bibinfo{author}{Tengg-Kobligk, H.V.}, \bibinfo{author}{Rohr, K.},
  \bibinfo{author}{Schnorr, C.}, \bibinfo{year}{2015}.
\newblock \bibinfo{title}{3d segmentation of vessels by incremental implicit
  polynomial fitting and convex optimization}, \bibinfo{publisher}{IEEE
  Computer Society}. pp. \bibinfo{pages}{1540--1543}.
\newblock \DOIprefix\doi{10.1109/ISBI.2015.7164171}.
\bibitem[{Biesdorf et~al.(2011)Biesdorf, Wörz, Müller, Weber, Heye, Hosch,
  Tengg-Kobligk and Rohr}]{Biesdorf_2011}
\bibinfo{author}{Biesdorf, A.}, \bibinfo{author}{Wörz, S.},
  \bibinfo{author}{Müller, T.}, \bibinfo{author}{Weber, T.F.},
  \bibinfo{author}{Heye, T.}, \bibinfo{author}{Hosch, W.},
  \bibinfo{author}{Tengg-Kobligk, H.V.}, \bibinfo{author}{Rohr, K.},
  \bibinfo{year}{2011}.
\newblock \bibinfo{title}{Model-based segmentation and motion analysis of the
  thoracic aorta from 4d ecg-gated cta images}, pp. \bibinfo{pages}{589--596}.
\newblock \DOIprefix\doi{10.1007/978-3-642-23623-5_74}.
\bibitem[{Bodur et~al.(2007)Bodur, Grady, Stillman, Setser, Funka-Lea and
  O'Donnell}]{Bodur_2007}
\bibinfo{author}{Bodur, O.}, \bibinfo{author}{Grady, L.},
  \bibinfo{author}{Stillman, A.}, \bibinfo{author}{Setser, R.},
  \bibinfo{author}{Funka-Lea, G.}, \bibinfo{author}{O'Donnell, T.},
  \bibinfo{year}{2007}.
\newblock \bibinfo{title}{{Semi-automatic aortic aneurysm analysis}}, in:
  \bibinfo{editor}{Manduca, A.}, \bibinfo{editor}{Hu, X.P.} (Eds.),
  \bibinfo{booktitle}{Medical Imaging 2007: Physiology, Function, and Structure
  from Medical Images}, \bibinfo{organization}{International Society for Optics
  and Photonics}. \bibinfo{publisher}{SPIE}. pp. \bibinfo{pages}{454 -- 463}.
\newblock \URLprefix \url{https://doi.org/10.1117/12.710719},
  \DOIprefix\doi{10.1117/12.710719}.
\bibitem[{Bolger(2008)}]{Bolger_2008}
\bibinfo{author}{Bolger, A.F.}, \bibinfo{year}{2008}.
\newblock \bibinfo{title}{Aortic intramural haematoma}.
\newblock \bibinfo{journal}{Heart} \bibinfo{volume}{94},
  \bibinfo{pages}{1670--1674}.
\newblock \DOIprefix\doi{10.1136/hrt.2007.132811}.
\bibitem[{Boskamp et~al.(2004)Boskamp, Rinck, Link, Kümmerlen, Stamm and
  Mildenberger}]{Boskamp_2004}
\bibinfo{author}{Boskamp, T.}, \bibinfo{author}{Rinck, D.},
  \bibinfo{author}{Link, F.}, \bibinfo{author}{Kümmerlen, B.},
  \bibinfo{author}{Stamm, G.}, \bibinfo{author}{Mildenberger, P.},
  \bibinfo{year}{2004}.
\newblock \bibinfo{title}{New vessel analysis tool for morphometric
  quantification and visualization of vessels in ct and mr imaging data sets}.
\newblock \bibinfo{journal}{Radiographics} \bibinfo{volume}{24},
  \bibinfo{pages}{287--297}.
\newblock \DOIprefix\doi{10.1148/rg.241035073}.
\bibitem[{Boykov et~al.(2001)Boykov, Veksler and Zabih}]{Boykov_2001}
\bibinfo{author}{Boykov, Y.}, \bibinfo{author}{Veksler, O.},
  \bibinfo{author}{Zabih, R.}, \bibinfo{year}{2001}.
\newblock \bibinfo{title}{Fast approximate energy minimization via graph cuts}.
\newblock \bibinfo{journal}{IEEE Transactions on Pattern Analysis and Machine
  Intelligence} , \bibinfo{pages}{1222 --
  1239}\DOIprefix\doi{10.1109/34.969114}.
\bibitem[{de~Bruijne et~al.(2002)de~Bruijne, Ginneken, Maintz, Niessen and
  Viergever}]{de_2002}
\bibinfo{author}{de~Bruijne, M.}, \bibinfo{author}{Ginneken, B.},
  \bibinfo{author}{Maintz, J.}, \bibinfo{author}{Niessen, W.},
  \bibinfo{author}{Viergever, M.}, \bibinfo{year}{2002}.
\newblock \bibinfo{title}{Active shape model based segmentation of abdominal
  aortic aneurysms in cta images}.
\newblock \bibinfo{journal}{Proceedings of SPIE - The International Society for
  Optical Engineering} \bibinfo{volume}{4684}.
\newblock \DOIprefix\doi{10.1117/12.467188}.
\bibitem[{de~Bruijne et~al.(2003)de~Bruijne, Ginneken, Niessen, Loog and
  Viergever}]{deBruijne_2003}
\bibinfo{author}{de~Bruijne, M.}, \bibinfo{author}{Ginneken, B.},
  \bibinfo{author}{Niessen, W.}, \bibinfo{author}{Loog, M.},
  \bibinfo{author}{Viergever, M.}, \bibinfo{year}{2003}.
\newblock \bibinfo{title}{Model-based segmentation of abdominal aortic
  aneurysms in cta images}.
\newblock \bibinfo{journal}{Proc SPIE} \bibinfo{volume}{5032}.
\newblock \DOIprefix\doi{10.1117/12.481367}.
\bibitem[{Bruyninckx et~al.(2010)Bruyninckx, Loeckx, Vandermeulen and
  Suetens}]{Bruyninckx_2010}
\bibinfo{author}{Bruyninckx, P.}, \bibinfo{author}{Loeckx, D.},
  \bibinfo{author}{Vandermeulen, D.}, \bibinfo{author}{Suetens, P.},
  \bibinfo{year}{2010}.
\newblock \bibinfo{title}{Segmentation of liver portal veins by global
  optimization}, \bibinfo{publisher}{SPIE}. p. \bibinfo{pages}{76241Z}.
\newblock \DOIprefix\doi{10.1117/12.843995}.
\bibitem[{Bustamante et~al.(2015)Bustamante, Petersson, Eriksson, Alehagen,
  Dyverfeldt, Carlhäll and Ebbers}]{Bustamante_2015}
\bibinfo{author}{Bustamante, M.}, \bibinfo{author}{Petersson, S.},
  \bibinfo{author}{Eriksson, J.}, \bibinfo{author}{Alehagen, U.},
  \bibinfo{author}{Dyverfeldt, P.}, \bibinfo{author}{Carlhäll, C.J.},
  \bibinfo{author}{Ebbers, T.}, \bibinfo{year}{2015}.
\newblock \bibinfo{title}{Atlas-based analysis of 4d flow cmr: Automated vessel
  segmentation and flow quantification}.
\newblock \bibinfo{journal}{Journal of Cardiovascular Magnetic Resonance}
  \bibinfo{volume}{17}.
\newblock \DOIprefix\doi{10.1186/s12968-015-0190-5}.
\bibitem[{Cao et~al.(2019)Cao, Shi, Ge, Xing, Zuo, Jia, Liu, He, Wang, Luan,
  Chai and Guo}]{Cao_2019}
\bibinfo{author}{Cao, L.}, \bibinfo{author}{Shi, R.}, \bibinfo{author}{Ge, Y.},
  \bibinfo{author}{Xing, L.}, \bibinfo{author}{Zuo, P.}, \bibinfo{author}{Jia,
  Y.}, \bibinfo{author}{Liu, J.}, \bibinfo{author}{He, Y.},
  \bibinfo{author}{Wang, X.}, \bibinfo{author}{Luan, S.},
  \bibinfo{author}{Chai, X.}, \bibinfo{author}{Guo, W.}, \bibinfo{year}{2019}.
\newblock \bibinfo{title}{Fully automatic segmentation of type b aortic
  dissection from cta images enabled by deep learning}.
\newblock \bibinfo{journal}{European Journal of Radiology}
  \bibinfo{volume}{121}.
\newblock \DOIprefix\doi{10.1016/j.ejrad.2019.108713}.
\bibitem[{Chen et~al.(2016)Chen, Mirebeau and Cohen}]{Chen_2016}
\bibinfo{author}{Chen, D.}, \bibinfo{author}{Mirebeau, J.M.},
  \bibinfo{author}{Cohen, L.D.}, \bibinfo{year}{2016}.
\newblock \bibinfo{title}{Vessel tree extraction using radius-lifted keypoints
  searching scheme and anisotropic fast marching method}.
\newblock \bibinfo{journal}{Journal of Algorithms and Computational Technology}
  \bibinfo{volume}{10}, \bibinfo{pages}{224--234}.
\newblock \DOIprefix\doi{10.1177/1748301816656289}.
\bibitem[{Chen et~al.(2021)Chen, Zhang, Mei, Liao, Xu, Li, Xiao, Guo, Zhang,
  Yan, Xiong and Ventikos}]{Chen_2021}
\bibinfo{author}{Chen, D.}, \bibinfo{author}{Zhang, X.}, \bibinfo{author}{Mei,
  Y.}, \bibinfo{author}{Liao, F.}, \bibinfo{author}{Xu, H.},
  \bibinfo{author}{Li, Z.}, \bibinfo{author}{Xiao, Q.}, \bibinfo{author}{Guo,
  W.}, \bibinfo{author}{Zhang, H.}, \bibinfo{author}{Yan, T.},
  \bibinfo{author}{Xiong, J.}, \bibinfo{author}{Ventikos, Y.},
  \bibinfo{year}{2021}.
\newblock \bibinfo{title}{Multi-stage learning for segmentation of aortic
  dissections using a prior aortic anatomy simplification}.
\newblock \bibinfo{journal}{Medical Image Analysis} \bibinfo{volume}{69}.
\newblock \DOIprefix\doi{10.1016/j.media.2020.101931}.
\bibitem[{Cheung et~al.(2021)Cheung, Bell, Nair, Menezies, Patel, Wan, Chou,
  Chen, Torii, Davies, Moon, Alexander and Jacob}]{Cheung_2021}
\bibinfo{author}{Cheung, W.K.}, \bibinfo{author}{Bell, R.},
  \bibinfo{author}{Nair, A.}, \bibinfo{author}{Menezies, L.},
  \bibinfo{author}{Patel, R.}, \bibinfo{author}{Wan, S.},
  \bibinfo{author}{Chou, K.}, \bibinfo{author}{Chen, J.},
  \bibinfo{author}{Torii, R.}, \bibinfo{author}{Davies, R.H.},
  \bibinfo{author}{Moon, J.C.}, \bibinfo{author}{Alexander, D.C.},
  \bibinfo{author}{Jacob, J.}, \bibinfo{year}{2021}.
\newblock \bibinfo{title}{A computationally efficient approach to segmentation
  of the aorta and coronary arteries using deep learning}.
\newblock \bibinfo{journal}{medRxiv}
  \DOIprefix\doi{10.1101/2021.02.18.21252005}.
\bibitem[{{\c{C}}i{\c{c}}ek et~al.(2016){\c{C}}i{\c{c}}ek, Abdulkadir,
  Lienkamp, Brox and Ronneberger}]{Cicek_2016}
\bibinfo{author}{{\c{C}}i{\c{c}}ek, {\"{O}}.}, \bibinfo{author}{Abdulkadir,
  A.}, \bibinfo{author}{Lienkamp, S.}, \bibinfo{author}{Brox, T.},
  \bibinfo{author}{Ronneberger, O.}, \bibinfo{year}{2016}.
\newblock \bibinfo{title}{3d u-net: Learning dense volumetric segmentation from
  sparse annotation}.
\newblock \bibinfo{journal}{International Conference on Medical Image Computing
  and Computer-Assisted Intervention} , \bibinfo{pages}{424--432}.
\bibitem[{Codari et~al.(2020)Codari, Pepe, Mistelbauer, Mastrodicasa, Walters,
  Willemink and Fleischmann}]{codari2020deep}
\bibinfo{author}{Codari, M.}, \bibinfo{author}{Pepe, A.},
  \bibinfo{author}{Mistelbauer, G.}, \bibinfo{author}{Mastrodicasa, D.},
  \bibinfo{author}{Walters, S.}, \bibinfo{author}{Willemink, M.J.},
  \bibinfo{author}{Fleischmann, D.}, \bibinfo{year}{2020}.
\newblock \bibinfo{title}{Deep reinforcement learning for localization of the
  aortic annulus in patients with aortic dissection}, in:
  \bibinfo{booktitle}{International Workshop on Thoracic Image Analysis},
  \bibinfo{organization}{Springer}. pp. \bibinfo{pages}{94--105}.
\bibitem[{Das et~al.(2006)Das, Mallya, Srikanth and Malladi}]{Das_2006}
\bibinfo{author}{Das, B.}, \bibinfo{author}{Mallya, Y.},
  \bibinfo{author}{Srikanth, S.}, \bibinfo{author}{Malladi, R.},
  \bibinfo{year}{2006}.
\newblock \bibinfo{title}{Aortic thrombus segmentation using narrow band active
  contour model}, in: \bibinfo{booktitle}{International Conference of the IEEE
  Engineering in Medicine and Biology Society}.
\newblock \DOIprefix\doi{10.1109/IEMBS.2006.4397423}.
\bibitem[{Du et~al.(2015)Du, Hu and Cai}]{Du_2015}
\bibinfo{author}{Du, T.}, \bibinfo{author}{Hu, D.}, \bibinfo{author}{Cai, D.},
  \bibinfo{year}{2015}.
\newblock \bibinfo{title}{Outflow boundary conditions for blood flow in
  arterial trees}.
\newblock \bibinfo{journal}{PLoS ONE} \bibinfo{volume}{10}.
\newblock \DOIprefix\doi{10.1371/journal.pone.0128597}.
\bibitem[{Duan et~al.(2016)Duan, Shi, Wang, Zhao and Chen}]{Duan_2016}
\bibinfo{author}{Duan, X.}, \bibinfo{author}{Shi, M.}, \bibinfo{author}{Wang,
  J.}, \bibinfo{author}{Zhao, H.}, \bibinfo{author}{Chen, D.},
  \bibinfo{year}{2016}.
\newblock \bibinfo{title}{Segmentation of the aortic dissection from \mbox{CT}
  images based on spatial continuity prior model}, in: \bibinfo{booktitle}{2016
  8th IEEE International Conference on Information Technology in Medicine and
  Education (ITME)}, pp. \bibinfo{pages}{275--280}.
\bibitem[{Duquette et~al.(2012)Duquette, Jodoin, Bouchot and
  Lalande}]{Duquette_2012}
\bibinfo{author}{Duquette, A.A.}, \bibinfo{author}{Jodoin, P.M.},
  \bibinfo{author}{Bouchot, O.}, \bibinfo{author}{Lalande, A.},
  \bibinfo{year}{2012}.
\newblock \bibinfo{title}{3d segmentation of abdominal aorta from ct-scan and
  mr images}.
\newblock \bibinfo{journal}{Computerized Medical Imaging and Graphics}
  \bibinfo{volume}{36}, \bibinfo{pages}{294--303}.
\newblock \DOIprefix\doi{10.1016/j.compmedimag.2011.12.001}.
\bibitem[{Egger et~al.(2009)Egger, Freisleben, Setser, Renapuraar, Biermann and
  O'Donnell}]{Egger_2009}
\bibinfo{author}{Egger, J.}, \bibinfo{author}{Freisleben, B.},
  \bibinfo{author}{Setser, R.}, \bibinfo{author}{Renapuraar, R.},
  \bibinfo{author}{Biermann, C.}, \bibinfo{author}{O'Donnell, T.},
  \bibinfo{year}{2009}.
\newblock \bibinfo{title}{Aorta segmentation for stent simulation}.
\newblock \bibinfo{journal}{MICCAI Workshop on Cardiovascular Interventional
  Imaging and Biophysical Modelling} .
\bibitem[{Egger et~al.(2012)Egger, Grosskopf, Nimsky, Kapur and
  Freisleben}]{Egger_2012}
\bibinfo{author}{Egger, J.}, \bibinfo{author}{Grosskopf, S.},
  \bibinfo{author}{Nimsky, C.}, \bibinfo{author}{Kapur, T.},
  \bibinfo{author}{Freisleben, B.}, \bibinfo{year}{2012}.
\newblock \bibinfo{title}{Modeling and visualization techniques for virtual
  stenting of aneurysms and stenoses}.
\newblock \bibinfo{journal}{Computerized Medical Imaging and Graphics}
  \bibinfo{volume}{36}, \bibinfo{pages}{183--203}.
\bibitem[{Egger et~al.(2007)Egger, Mostarki, Grobkopf and
  Freisleben}]{Egger_2007_preoperative}
\bibinfo{author}{Egger, J.}, \bibinfo{author}{Mostarki, Z.},
  \bibinfo{author}{Grobkopf, S.}, \bibinfo{author}{Freisleben, B.},
  \bibinfo{year}{2007}.
\newblock \bibinfo{title}{Preoperative measurement of aneurysms and stenosis
  and stentsimulation for endovascular treatment}, in: \bibinfo{booktitle}{2007
  4th IEEE International Symposium on Biomedical Imaging: From Nano to Macro},
  pp. \bibinfo{pages}{392--395}.
\newblock \DOIprefix\doi{10.1109/ISBI.2007.356871}.
\bibitem[{{Egger} et~al.(2007){Egger}, {Mostarkic}, {Grosskopf} and
  {Freisleben}}]{Egger_2007}
\bibinfo{author}{{Egger}, J.}, \bibinfo{author}{{Mostarkic}, Z.},
  \bibinfo{author}{{Grosskopf}, S.}, \bibinfo{author}{{Freisleben}, B.},
  \bibinfo{year}{2007}.
\newblock \bibinfo{title}{A fast vessel centerline extraction algorithm for
  catheter simulation}, in: \bibinfo{booktitle}{Twentieth IEEE International
  Symposium on Computer-Based Medical Systems (CBMS'07)}, pp.
  \bibinfo{pages}{177--182}.
\newblock \DOIprefix\doi{10.1109/CBMS.2007.5}.
\bibitem[{Erbel and Aboyans(2014)}]{Erbel_2014}
\bibinfo{author}{Erbel, R.}, \bibinfo{author}{Aboyans, V.e.a.},
  \bibinfo{year}{2014}.
\newblock \bibinfo{title}{2014 esc guidelines on the diagnosis and treatment of
  aortic diseases}.
\newblock \bibinfo{journal}{European Heart Journal} \bibinfo{volume}{35},
  \bibinfo{pages}{2873--2926\colrefE}.
\bibitem[{Erhart et~al.(2015)Erhart, Hyhlik-Dürr, Geisbüsch, Kotelis,
  Müller-Eschner, Gasser, {von Tengg-Kobligk} and Böckler}]{Erhart_2014}
\bibinfo{author}{Erhart, P.}, \bibinfo{author}{Hyhlik-Dürr, A.},
  \bibinfo{author}{Geisbüsch, P.}, \bibinfo{author}{Kotelis, D.},
  \bibinfo{author}{Müller-Eschner, M.}, \bibinfo{author}{Gasser, T.},
  \bibinfo{author}{{von Tengg-Kobligk}, H.}, \bibinfo{author}{Böckler, D.},
  \bibinfo{year}{2015}.
\newblock \bibinfo{title}{Finite element analysis in asymptomatic, symptomatic,
  and ruptured abdominal aortic aneurysms: In search of new rupture risk
  predictors}.
\newblock \bibinfo{journal}{European Journal of Vascular and Endovascular
  Surgery} \bibinfo{volume}{49}, \bibinfo{pages}{239--245}.
\newblock \URLprefix
  \url{https://www.sciencedirect.com/science/article/pii/S1078588414006376},
  \DOIprefix\doi{https://doi.org/10.1016/j.ejvs.2014.11.010}.
\bibitem[{Fantazzini et~al.(2020)Fantazzini, Esposito, Finotello, Auricchio,
  Pane, Basso, Spinella and Conti}]{Fantazzini_2020}
\bibinfo{author}{Fantazzini, A.}, \bibinfo{author}{Esposito, M.},
  \bibinfo{author}{Finotello, A.}, \bibinfo{author}{Auricchio, F.},
  \bibinfo{author}{Pane, B.}, \bibinfo{author}{Basso, C.},
  \bibinfo{author}{Spinella, G.}, \bibinfo{author}{Conti, M.},
  \bibinfo{year}{2020}.
\newblock \bibinfo{title}{3d automatic segmentation of aortic computed
  tomography angiography combining multi-view 2d convolutional neural
  networks}.
\newblock \bibinfo{journal}{Cardiovascular Engineering and Technology}
  \bibinfo{volume}{11}, \bibinfo{pages}{576--586}.
\newblock \DOIprefix\doi{10.1007/s13239-020-00481-z}.
\bibitem[{Frangi et~al.(1998)Frangi, Niessen, Vincken and
  Viergever}]{Frangi_1998}
\bibinfo{author}{Frangi, A.F.}, \bibinfo{author}{Niessen, W.J.},
  \bibinfo{author}{Vincken, K.L.}, \bibinfo{author}{Viergever, M.A.},
  \bibinfo{year}{1998}.
\newblock \bibinfo{title}{Multiscale vessel enhancement filtering}, in:
  \bibinfo{editor}{Wells, W.M.}, \bibinfo{editor}{Colchester, A.},
  \bibinfo{editor}{Delp, S.} (Eds.), \bibinfo{booktitle}{Medical Image
  Computing and Computer-Assisted Intervention --- MICCAI'98},
  \bibinfo{publisher}{Springer Berlin Heidelberg}, \bibinfo{address}{Berlin,
  Heidelberg}. pp. \bibinfo{pages}{130--137}.
\bibitem[{Freiman et~al.(2012)Freiman, Joskowicz, Broide, Natanzon, Nammer,
  Shilon, Weizman and Sosna}]{Freiman_2012}
\bibinfo{author}{Freiman, M.}, \bibinfo{author}{Joskowicz, L.},
  \bibinfo{author}{Broide, N.}, \bibinfo{author}{Natanzon, M.},
  \bibinfo{author}{Nammer, E.}, \bibinfo{author}{Shilon, O.},
  \bibinfo{author}{Weizman, L.}, \bibinfo{author}{Sosna, J.},
  \bibinfo{year}{2012}.
\newblock \bibinfo{title}{Carotid vasculature modeling from patient ct
  angiography studies for interventional procedures simulation}.
\newblock \bibinfo{journal}{International Journal of Computer Assisted
  Radiology and Surgery} \bibinfo{volume}{7}, \bibinfo{pages}{799--812}.
\newblock \DOIprefix\doi{10.1007/s11548-012-0673-x}.
\bibitem[{Gallagher and Raff(2008)}]{gallagher2008use}
\bibinfo{author}{Gallagher, M.J.}, \bibinfo{author}{Raff, G.L.},
  \bibinfo{year}{2008}.
\newblock \bibinfo{title}{Use of multislice ct for the evaluation of emergency
  room patients with chest pain: the so-called “triple rule-out”}.
\newblock \bibinfo{journal}{Catheterization and cardiovascular interventions}
  \bibinfo{volume}{71}, \bibinfo{pages}{92--99}.
\bibitem[{Gamechi et~al.(2019)Gamechi, Bons, Giordano, Bos, Budde, Kofoed,
  Pedersen, Roos-Hesselink and de~Bruijne}]{Gamechi_2019}
\bibinfo{author}{Gamechi, Z.S.}, \bibinfo{author}{Bons, L.R.},
  \bibinfo{author}{Giordano, M.}, \bibinfo{author}{Bos, D.},
  \bibinfo{author}{Budde, R.P.}, \bibinfo{author}{Kofoed, K.F.},
  \bibinfo{author}{Pedersen, J.H.}, \bibinfo{author}{Roos-Hesselink, J.W.},
  \bibinfo{author}{de~Bruijne, M.}, \bibinfo{year}{2019}.
\newblock \bibinfo{title}{Automated 3d segmentation and diameter measurement of
  the thoracic aorta on non-contrast enhanced ct}.
\newblock \bibinfo{journal}{European Radiology} \bibinfo{volume}{29},
  \bibinfo{pages}{4613--4623}.
\newblock \DOIprefix\doi{10.1007/s00330-018-5931-z}.
\bibitem[{Gao et~al.(2016)Gao, Kitslaar, Budde, Tu, de~Graaf, Xu, Xu, Scholte,
  Dijkstra and Reiber}]{Gao_2016}
\bibinfo{author}{Gao, X.}, \bibinfo{author}{Kitslaar, P.H.},
  \bibinfo{author}{Budde, R.P.}, \bibinfo{author}{Tu, S.},
  \bibinfo{author}{de~Graaf, M.A.}, \bibinfo{author}{Xu, L.},
  \bibinfo{author}{Xu, B.}, \bibinfo{author}{Scholte, A.J.},
  \bibinfo{author}{Dijkstra, J.}, \bibinfo{author}{Reiber, J.H.},
  \bibinfo{year}{2016}.
\newblock \bibinfo{title}{Automatic detection of aorto-femoral vessel
  trajectory from whole-body computed tomography angiography data sets}.
\newblock \bibinfo{journal}{International Journal of Cardiovascular Imaging}
  \bibinfo{volume}{32}, \bibinfo{pages}{1311--1322}.
\newblock \DOIprefix\doi{10.1007/s10554-016-0901-5}.
\bibitem[{Hager et~al.(2007)Hager, Kanz, Kaemmerer, Schreiber and
  Hess}]{Hager_2007}
\bibinfo{author}{Hager, A.}, \bibinfo{author}{Kanz, S.},
  \bibinfo{author}{Kaemmerer, H.}, \bibinfo{author}{Schreiber, C.},
  \bibinfo{author}{Hess, J.}, \bibinfo{year}{2007}.
\newblock \bibinfo{title}{Coarctation long-term assessment (coala):
  Significance of arterial hypertension in a cohort of 404 patients up to 27
  years after surgical repair of isolated coarctation of the aorta, even in the
  absence of restenosis and prosthetic material}.
\newblock \bibinfo{journal}{Journal of Thoracic and Cardiovascular Surgery}
  \bibinfo{volume}{134}.
\newblock \DOIprefix\doi{10.1016/j.jtcvs.2007.04.027}.
\bibitem[{Hahn et~al.(2020)Hahn, Mistelbauer, Higashigaito, Koci, Willemink,
  Sailer, Fischbein and Fleischmann}]{Hahn_2020}
\bibinfo{author}{Hahn, L.}, \bibinfo{author}{Mistelbauer, G.},
  \bibinfo{author}{Higashigaito, K.}, \bibinfo{author}{Koci, M.},
  \bibinfo{author}{Willemink, M.}, \bibinfo{author}{Sailer, A.},
  \bibinfo{author}{Fischbein, M.}, \bibinfo{author}{Fleischmann, D.},
  \bibinfo{year}{2020}.
\newblock \bibinfo{title}{Ct-based true- and false-lumen segmentation in type b
  aortic dissection using machine learning}.
\newblock \bibinfo{journal}{Radiology: Cardiothoracic Imaging}
  \bibinfo{volume}{2}, \bibinfo{pages}{e190179}.
\newblock \DOIprefix\doi{10.1148/ryct.2020190179}.
\bibitem[{Harris et~al.(2012)Harris, Braverman, Eagle, Woznicki, Pyeritz,
  Myrmel, Peterson, Voehringer, Fattori, Januzzi, Gilon, Montgomery, Nienaber,
  Trimarchi, Isselbacher and Evangelista}]{Harris_2012}
\bibinfo{author}{Harris, K.M.}, \bibinfo{author}{Braverman, A.C.},
  \bibinfo{author}{Eagle, K.A.}, \bibinfo{author}{Woznicki, E.M.},
  \bibinfo{author}{Pyeritz, R.E.}, \bibinfo{author}{Myrmel, T.},
  \bibinfo{author}{Peterson, M.D.}, \bibinfo{author}{Voehringer, M.},
  \bibinfo{author}{Fattori, R.}, \bibinfo{author}{Januzzi, J.L.},
  \bibinfo{author}{Gilon, D.}, \bibinfo{author}{Montgomery, D.G.},
  \bibinfo{author}{Nienaber, C.A.}, \bibinfo{author}{Trimarchi, S.},
  \bibinfo{author}{Isselbacher, E.M.}, \bibinfo{author}{Evangelista, A.},
  \bibinfo{year}{2012}.
\newblock \bibinfo{title}{Acute aortic intramural hematoma: An analysis from
  the international registry of acute aortic dissection}.
\newblock \bibinfo{journal}{Circulation} \bibinfo{volume}{126}.
\newblock \DOIprefix\doi{10.1161/CIRCULATIONAHA.111.084541}.
\bibitem[{{Heimann} et~al.(2009){Heimann}, {van Ginneken}, {Styner},
  {Arzhaeva}, {Aurich}, {Bauer}, {Beck}, {Becker}, {Beichel}, {Bekes}, {Bello},
  {Binnig}, {Bischof}, {Bornik}, {Cashman}, {Chi}, {Cordova}, {Dawant},
  {Fidrich}, {Furst}, {Furukawa}, {Grenacher}, {Hornegger}, {Kainmueller},
  {Kitney}, {Kobatake}, {Lamecker}, {Lange}, {Lee}, {Lennon}, {Li}, {Li},
  {Meinzer}, {Nemeth}, {Raicu}, {Rau}, {van Rikxoort}, {Rousson}, {Rusko},
  {Saddi}, {Schmidt}, {Seghers}, {Shimizu}, {Slagmolen}, {Sorantin}, {Soza},
  {Susomboon}, {Waite}, {Wimmer} and {Wolf}}]{Heimann_2009}
\bibinfo{author}{{Heimann}, T.}, \bibinfo{author}{{van Ginneken}, B.},
  \bibinfo{author}{{Styner}, M.A.}, \bibinfo{author}{{Arzhaeva}, Y.},
  \bibinfo{author}{{Aurich}, V.}, \bibinfo{author}{{Bauer}, C.},
  \bibinfo{author}{{Beck}, A.}, \bibinfo{author}{{Becker}, C.},
  \bibinfo{author}{{Beichel}, R.}, \bibinfo{author}{{Bekes}, G.},
  \bibinfo{author}{{Bello}, F.}, \bibinfo{author}{{Binnig}, G.},
  \bibinfo{author}{{Bischof}, H.}, \bibinfo{author}{{Bornik}, A.},
  \bibinfo{author}{{Cashman}, P.M.M.}, \bibinfo{author}{{Chi}, Y.},
  \bibinfo{author}{{Cordova}, A.}, \bibinfo{author}{{Dawant}, B.M.},
  \bibinfo{author}{{Fidrich}, M.}, \bibinfo{author}{{Furst}, J.D.},
  \bibinfo{author}{{Furukawa}, D.}, \bibinfo{author}{{Grenacher}, L.},
  \bibinfo{author}{{Hornegger}, J.}, \bibinfo{author}{{Kainmueller}, D.},
  \bibinfo{author}{{Kitney}, R.I.}, \bibinfo{author}{{Kobatake}, H.},
  \bibinfo{author}{{Lamecker}, H.}, \bibinfo{author}{{Lange}, T.},
  \bibinfo{author}{{Lee}, J.}, \bibinfo{author}{{Lennon}, B.},
  \bibinfo{author}{{Li}, R.}, \bibinfo{author}{{Li}, S.},
  \bibinfo{author}{{Meinzer}, H.}, \bibinfo{author}{{Nemeth}, G.},
  \bibinfo{author}{{Raicu}, D.S.}, \bibinfo{author}{{Rau}, A.},
  \bibinfo{author}{{van Rikxoort}, E.M.}, \bibinfo{author}{{Rousson}, M.},
  \bibinfo{author}{{Rusko}, L.}, \bibinfo{author}{{Saddi}, K.A.},
  \bibinfo{author}{{Schmidt}, G.}, \bibinfo{author}{{Seghers}, D.},
  \bibinfo{author}{{Shimizu}, A.}, \bibinfo{author}{{Slagmolen}, P.},
  \bibinfo{author}{{Sorantin}, E.}, \bibinfo{author}{{Soza}, G.},
  \bibinfo{author}{{Susomboon}, R.}, \bibinfo{author}{{Waite}, J.M.},
  \bibinfo{author}{{Wimmer}, A.}, \bibinfo{author}{{Wolf}, I.},
  \bibinfo{year}{2009}.
\newblock \bibinfo{title}{Comparison and evaluation of methods for liver
  segmentation from ct datasets}.
\newblock \bibinfo{journal}{IEEE Transactions on Medical Imaging}
  \bibinfo{volume}{28}, \bibinfo{pages}{1251--1265}.
\newblock \DOIprefix\doi{10.1109/TMI.2009.2013851}.
\bibitem[{Hepp et~al.(2020)Hepp, Fischer, Winkelmann, Baldenhofer, Kuestner,
  Nikolaou, Yang and Gatidis}]{Hepp_2020}
\bibinfo{author}{Hepp, T.}, \bibinfo{author}{Fischer, M.},
  \bibinfo{author}{Winkelmann, M.T.}, \bibinfo{author}{Baldenhofer, S.},
  \bibinfo{author}{Kuestner, T.}, \bibinfo{author}{Nikolaou, K.},
  \bibinfo{author}{Yang, B.}, \bibinfo{author}{Gatidis, S.},
  \bibinfo{year}{2020}.
\newblock \bibinfo{title}{Fully automated segmentation and shape analysis of
  the thoracic aorta in non-contrast-enhanced magnetic resonance images of the
  german national cohort study}.
\newblock \bibinfo{journal}{J Thorac Imaging}
  \DOIprefix\doi{10.1097/RTI.0000000000000522}.
\bibitem[{Herment et~al.(2010)Herment, Kachenoura, Lefort, Bensalah, Dogui,
  Frouin, Mousseaux and Cesare}]{Herment_2010}
\bibinfo{author}{Herment, A.}, \bibinfo{author}{Kachenoura, N.},
  \bibinfo{author}{Lefort, M.}, \bibinfo{author}{Bensalah, M.},
  \bibinfo{author}{Dogui, A.}, \bibinfo{author}{Frouin, F.},
  \bibinfo{author}{Mousseaux, E.}, \bibinfo{author}{Cesare, A.D.},
  \bibinfo{year}{2010}.
\newblock \bibinfo{title}{Automated segmentation of the aorta from phase
  contrast mr images: Validation against expert tracing in healthy volunteers
  and in patients with a dilated aorta}.
\newblock \bibinfo{journal}{Journal of Magnetic Resonance Imaging}
  \bibinfo{volume}{31}, \bibinfo{pages}{881--888}.
\newblock \DOIprefix\doi{10.1002/jmri.22124}.
\bibitem[{Hesamian et~al.(2019)Hesamian, Jia, He and Kennedy}]{Hesamian_2019}
\bibinfo{author}{Hesamian, M.}, \bibinfo{author}{Jia, W.}, \bibinfo{author}{He,
  X.}, \bibinfo{author}{Kennedy, P.}, \bibinfo{year}{2019}.
\newblock \bibinfo{title}{Deep learning techniques for medical image
  segmentation: Achievements and challenges}.
\newblock \bibinfo{journal}{Journal of Digital Imaging} ,
  \bibinfo{pages}{582--596}.
\bibitem[{Howard et~al.(2020)Howard, Zaman, Ragavan, Hall, Leonard, Sutanto,
  Ramadoss, Razvi, Linton, Bharath, Shun-Shin, Rueckert, Francis and
  Cole}]{Howard_2020}
\bibinfo{author}{Howard, J.P.}, \bibinfo{author}{Zaman, S.},
  \bibinfo{author}{Ragavan, A.}, \bibinfo{author}{Hall, K.},
  \bibinfo{author}{Leonard, G.}, \bibinfo{author}{Sutanto, S.},
  \bibinfo{author}{Ramadoss, V.}, \bibinfo{author}{Razvi, Y.},
  \bibinfo{author}{Linton, N.F.}, \bibinfo{author}{Bharath, A.},
  \bibinfo{author}{Shun-Shin, M.}, \bibinfo{author}{Rueckert, D.},
  \bibinfo{author}{Francis, D.}, \bibinfo{author}{Cole, G.},
  \bibinfo{year}{2020}.
\newblock \bibinfo{title}{Automated analysis and detection of abnormalities in
  transaxial anatomical cardiovascular magnetic resonance images: a proof of
  concept study with potential to optimize image acquisition}.
\newblock \bibinfo{journal}{The International Journal of Cardiovascular
  Imaging} \URLprefix
  \url{http://link.springer.com/10.1007/s10554-020-02050-w},
  \DOIprefix\doi{10.1007/s10554-020-02050-w}.
\bibitem[{Jin et~al.(2021)Jin, Pepe, Li, Gsaxner and Egger}]{Jin_2021}
\bibinfo{author}{Jin, Y.}, \bibinfo{author}{Pepe, A.}, \bibinfo{author}{Li,
  J.}, \bibinfo{author}{Gsaxner, C.}, \bibinfo{author}{Egger, J.},
  \bibinfo{year}{2021}.
\newblock \bibinfo{title}{Deep learning and particle filter-based aortic
  dissection vessel tree segmentation}, in: \bibinfo{booktitle}{Medical Imaging
  2021: Biomedical Applications in Molecular, Structural, and Functional
  Imaging}, \bibinfo{organization}{International Society for Optics and
  Photonics}. p.~\bibinfo{pages}{6}.
\bibitem[{Kass et~al.(1988)Kass, Witkin and Terzopoulos}]{kass1988snakes}
\bibinfo{author}{Kass, M.}, \bibinfo{author}{Witkin, A.},
  \bibinfo{author}{Terzopoulos, D.}, \bibinfo{year}{1988}.
\newblock \bibinfo{title}{Snakes: Active contour models}.
\newblock \bibinfo{journal}{International journal of computer vision}
  \bibinfo{volume}{1}, \bibinfo{pages}{321--331}.
\bibitem[{Khan and Nair(2002)}]{Khan_2002}
\bibinfo{author}{Khan, I.A.}, \bibinfo{author}{Nair, C.K.},
  \bibinfo{year}{2002}.
\newblock \bibinfo{title}{Clinical, diagnostic, and management perspectives of
  aortic dissection}.
\newblock \bibinfo{journal}{Chest} \bibinfo{volume}{122 (1)},
  \bibinfo{pages}{311--328}.
\bibitem[{Kim et~al.(2007)Kim, Seol, Choi, Oh, Kim and Sun}]{kim_2007}
\bibinfo{author}{Kim, H.C.}, \bibinfo{author}{Seol, Y.}, \bibinfo{author}{Choi,
  S.}, \bibinfo{author}{Oh, J.}, \bibinfo{author}{Kim, M.},
  \bibinfo{author}{Sun, K.}, \bibinfo{year}{2007}.
\newblock \bibinfo{title}{A study of aaa image segmentation technique using
  geometric active contour model with morphological gradient edge function}.
\newblock \bibinfo{journal}{Conference proceedings : ... Annual International
  Conference of the IEEE Engineering in Medicine and Biology Society. IEEE
  Engineering in Medicine and Biology Society. Conference}
  \bibinfo{volume}{2007}, \bibinfo{pages}{4437--40}.
\newblock \DOIprefix\doi{10.1109/IEMBS.2007.4353323}.
\bibitem[{Kirişli(2010)}]{Kirili_2010}
\bibinfo{author}{Kirişli, H.e.a.}, \bibinfo{year}{2010}.
\newblock \bibinfo{title}{Evaluation of a multi‐atlas based method for
  segmentation of cardiac cta data: a large‐scale, multicenter, and
  multivendor study}.
\newblock \bibinfo{journal}{Medical Physics} \DOIprefix\doi{10.1118/1.3512795}.
\bibitem[{Kosasih(2020)}]{Kosasih_2020}
\bibinfo{author}{Kosasih, R.}, \bibinfo{year}{2020}.
\newblock \bibinfo{title}{Automatic segmentation of abdominal aortic aneurism
  (aaa) by using active contour models}.
\newblock \bibinfo{journal}{Scientific Journal of Informatics}
  \bibinfo{volume}{7}, \bibinfo{pages}{2407--7658}.
\newblock \URLprefix \url{http://journal.unnes.ac.id/nju/index.php/sji}.
\bibitem[{Koutouzi et~al.(2017)Koutouzi, Sandstr{\"o}m, Skoog, Roos and
  Falkenberg}]{koutouzi20173d}
\bibinfo{author}{Koutouzi, G.}, \bibinfo{author}{Sandstr{\"o}m, C.},
  \bibinfo{author}{Skoog, P.}, \bibinfo{author}{Roos, H.},
  \bibinfo{author}{Falkenberg, M.}, \bibinfo{year}{2017}.
\newblock \bibinfo{title}{3d image fusion to localise intercostal arteries
  during tevar}.
\newblock \bibinfo{journal}{EJVES short reports} \bibinfo{volume}{35},
  \bibinfo{pages}{7--10}.
\bibitem[{Krissian et~al.(2014)Krissian, Carreira, Esclarin and
  Maynar}]{Krissian_2014}
\bibinfo{author}{Krissian, K.}, \bibinfo{author}{Carreira, J.M.},
  \bibinfo{author}{Esclarin, J.}, \bibinfo{author}{Maynar, M.},
  \bibinfo{year}{2014}.
\newblock \bibinfo{title}{Semi-automatic segmentation and detection of aorta
  dissection wall in \mbox{MDCT} angiography}.
\newblock \bibinfo{journal}{Medical Image Analysis} \bibinfo{volume}{18 (1)},
  \bibinfo{pages}{83 --102}.
\bibitem[{Krissian et~al.(2000)Krissian, Malandain, Ayache, Vaillant and
  Trousset}]{Krissian_2000}
\bibinfo{author}{Krissian, K.}, \bibinfo{author}{Malandain, G.},
  \bibinfo{author}{Ayache, N.}, \bibinfo{author}{Vaillant, R.},
  \bibinfo{author}{Trousset, Y.}, \bibinfo{year}{2000}.
\newblock \bibinfo{title}{Model-based detection of tubular structures in 3d
  images}.
\newblock \bibinfo{journal}{Computer vision and image understanding}
  \bibinfo{volume}{80}, \bibinfo{pages}{130--171}.
\bibitem[{Kurugol et~al.(2015)Kurugol, Come, Diaz, Ross, Kinney, Black-Shinn,
  Hokanson, Budoff, Washko and Estepar}]{Kurugol2015}
\bibinfo{author}{Kurugol, S.}, \bibinfo{author}{Come, C.E.},
  \bibinfo{author}{Diaz, A.A.}, \bibinfo{author}{Ross, J.C.},
  \bibinfo{author}{Kinney, G.L.}, \bibinfo{author}{Black-Shinn, J.L.},
  \bibinfo{author}{Hokanson, J.E.}, \bibinfo{author}{Budoff, M.J.},
  \bibinfo{author}{Washko, G.R.}, \bibinfo{author}{Estepar, R.S.J.},
  \bibinfo{year}{2015}.
\newblock \bibinfo{title}{Automated quantitative 3d analysis of aorta size,
  morphology, and mural calcification distributions}.
\newblock \bibinfo{journal}{Medical Physics} \bibinfo{volume}{42},
  \bibinfo{pages}{5467--5478}.
\newblock \DOIprefix\doi{10.1118/1.4924500}.
\bibitem[{Lambert et~al.(2020)Lambert, Petitjean, Dubray and
  Kuan}]{lambert2020segthor}
\bibinfo{author}{Lambert, Z.}, \bibinfo{author}{Petitjean, C.},
  \bibinfo{author}{Dubray, B.}, \bibinfo{author}{Kuan, S.},
  \bibinfo{year}{2020}.
\newblock \bibinfo{title}{Segthor: Segmentation of thoracic organs at risk in
  ct images}, in: \bibinfo{booktitle}{2020 Tenth International Conference on
  Image Processing Theory, Tools and Applications (IPTA)},
  \bibinfo{organization}{IEEE}. pp. \bibinfo{pages}{1--6}.
\bibitem[{Lareyre et~al.(2019)Lareyre, Adam, Carrier, Dommerc, Mialhe and
  Raffort}]{Lareyre_2019}
\bibinfo{author}{Lareyre, F.}, \bibinfo{author}{Adam, C.},
  \bibinfo{author}{Carrier, M.}, \bibinfo{author}{Dommerc, C.},
  \bibinfo{author}{Mialhe, C.}, \bibinfo{author}{Raffort, J.},
  \bibinfo{year}{2019}.
\newblock \bibinfo{title}{A fully automated pipeline for mining abdominal
  aortic aneurysm using image segmentation}.
\newblock \bibinfo{journal}{Scientific Reports} \bibinfo{volume}{9}.
\newblock \DOIprefix\doi{10.1038/s41598-019-50251-8}.
\bibitem[{Lawonn et~al.(2018)Lawonn, Smit, B{\"u}hler and
  Preim}]{lawonn2018survey}
\bibinfo{author}{Lawonn, K.}, \bibinfo{author}{Smit, N.N.},
  \bibinfo{author}{B{\"u}hler, K.}, \bibinfo{author}{Preim, B.},
  \bibinfo{year}{2018}.
\newblock \bibinfo{title}{A survey on multimodal medical data visualization},
  in: \bibinfo{booktitle}{Computer Graphics Forum},
  \bibinfo{organization}{Wiley Online Library}. pp. \bibinfo{pages}{413--438}.
\bibitem[{Lee et~al.(2017)Lee, Kang and Lee}]{Lee_2017}
\bibinfo{author}{Lee, S.H.}, \bibinfo{author}{Kang, J.}, \bibinfo{author}{Lee,
  S.}, \bibinfo{year}{2017}.
\newblock \bibinfo{title}{Enhanced particle-filtering framework for vessel
  segmentation and tracking}.
\newblock \bibinfo{journal}{Computer Methods and Programs in Biomedicine}
  \bibinfo{volume}{148}, \bibinfo{pages}{99--112}.
\newblock \DOIprefix\doi{10.1016/j.cmpb.2017.06.017}.
\bibitem[{LeMaire and Russel(2011)}]{LeMaire_2011}
\bibinfo{author}{LeMaire, S.A.}, \bibinfo{author}{Russel, L.},
  \bibinfo{year}{2011}.
\newblock \bibinfo{title}{Epidemiology of thoracic aortic dissection}.
\newblock \bibinfo{journal}{Nature Reviews Cardiology} \bibinfo{volume}{8},
  \bibinfo{pages}{103--113}.
\bibitem[{Levy et~al.(2022)Levy, Goyal, Grigorova, Farci and
  Le}]{levy2022aortic}
\bibinfo{author}{Levy, D.}, \bibinfo{author}{Goyal, A.},
  \bibinfo{author}{Grigorova, Y.}, \bibinfo{author}{Farci, F.},
  \bibinfo{author}{Le, J.K.}, \bibinfo{year}{2022}.
\newblock \bibinfo{title}{Aortic dissection}, in:
  \bibinfo{booktitle}{StatPearls [Internet]}. \bibinfo{publisher}{StatPearls
  publishing}.
\bibitem[{Li et~al.(2018)Li, Cao, Cheng, Bowen and Guo}]{Jianning}
\bibinfo{author}{Li, J.}, \bibinfo{author}{Cao, L.}, \bibinfo{author}{Cheng,
  W.}, \bibinfo{author}{Bowen, M.}, \bibinfo{author}{Guo, W.},
  \bibinfo{year}{2018}.
\newblock \bibinfo{title}{Towards automatic measurement of type \mbox{B} aortic
  dissection parameters: Methods, applications and perspective}, in:
  \bibinfo{booktitle}{Intravascular Imaging and Computer Assisted Stenting and
  Large-Scale Annotation of Biomedical Data and Expert Label Synthesis},
  \bibinfo{publisher}{Springer International Publishing},
  \bibinfo{address}{Cham}. pp. \bibinfo{pages}{64--72}.
\bibitem[{Li et~al.(2006)Li, Wu, Chen and Sonka}]{Li_2006}
\bibinfo{author}{Li, K.}, \bibinfo{author}{Wu, X.}, \bibinfo{author}{Chen,
  D.Z.}, \bibinfo{author}{Sonka, M.}, \bibinfo{year}{2006}.
\newblock \bibinfo{title}{Optimal surface segmentation in volumetric images-a
  graph-theoretic approach}.
\newblock \bibinfo{journal}{IEEE Transactions on Pattern Analysis and Machine
  Intelligence} \bibinfo{volume}{28}, \bibinfo{pages}{119--134}.
\newblock \DOIprefix\doi{10.1109/TPAMI.2006.19}.
\bibitem[{Li et~al.(2019)Li, Feng, Feng, An, Gao, Lu and Zhou}]{Li_2019}
\bibinfo{author}{Li, Z.}, \bibinfo{author}{Feng, J.}, \bibinfo{author}{Feng,
  Z.}, \bibinfo{author}{An, Y.}, \bibinfo{author}{Gao, Y.},
  \bibinfo{author}{Lu, B.}, \bibinfo{author}{Zhou, J.}, \bibinfo{year}{2019}.
\newblock \bibinfo{title}{Lumen segmentation of aortic dissection with cascaded
  convolutional network}, \bibinfo{publisher}{Springer Verlag}. pp.
  \bibinfo{pages}{122--130}.
\newblock \DOIprefix\doi{10.1007/978-3-030-12029-0_14}.
\bibitem[{Litjens et~al.(2017)Litjens, Kooi, Bejnordi, Setio, Ciompi,
  Ghafoorian, van~der Laak, van Ginneken and Sánchez}]{Litjens_2017}
\bibinfo{author}{Litjens, G.}, \bibinfo{author}{Kooi, T.},
  \bibinfo{author}{Bejnordi, B.E.}, \bibinfo{author}{Setio, A.A.A.},
  \bibinfo{author}{Ciompi, F.}, \bibinfo{author}{Ghafoorian, M.},
  \bibinfo{author}{van~der Laak, J.A.W.M.}, \bibinfo{author}{van Ginneken, B.},
  \bibinfo{author}{Sánchez, C.I.}, \bibinfo{year}{2017}.
\newblock \bibinfo{title}{A survey on deep learning in medical image analysis}
  \URLprefix \url{http://arxiv.org/abs/1702.05747
  http://dx.doi.org/10.1016/j.media.2017.07.005},
  \DOIprefix\doi{10.1016/j.media.2017.07.005}.
\bibitem[{Loncaric et~al.(2000)Loncaric, Subasic and Sorantin}]{Loncaric_2000}
\bibinfo{author}{Loncaric, S.}, \bibinfo{author}{Subasic, M.},
  \bibinfo{author}{Sorantin, E.}, \bibinfo{year}{2000}.
\newblock \bibinfo{title}{3-d deformable model for abdominal aortic aneurysm
  segmentation from ct images}, pp. \bibinfo{pages}{139 -- 144}.
\newblock \DOIprefix\doi{10.1109/ISPA.2000.914904}.
\bibitem[{Long et~al.(2014)Long, Shelhamer and Darrell}]{Long_2014}
\bibinfo{author}{Long, J.}, \bibinfo{author}{Shelhamer, E.},
  \bibinfo{author}{Darrell, T.}, \bibinfo{year}{2014}.
\newblock \bibinfo{title}{Fully convolutional networks for semantic
  segmentation}.
\newblock \bibinfo{journal}{CoRR} \bibinfo{volume}{abs/1411.4038}.
\newblock \URLprefix \url{http://arxiv.org/abs/1411.4038}.
\bibitem[{López-Linares et~al.(2018)López-Linares, Aranjuelo, Kabongo,
  Maclair, Lete, Ceresa, García-Familiar, Macía and Ballester}]{Linares_2018}
\bibinfo{author}{López-Linares, K.}, \bibinfo{author}{Aranjuelo, N.},
  \bibinfo{author}{Kabongo, L.}, \bibinfo{author}{Maclair, G.},
  \bibinfo{author}{Lete, N.}, \bibinfo{author}{Ceresa, M.},
  \bibinfo{author}{García-Familiar, A.}, \bibinfo{author}{Macía, I.},
  \bibinfo{author}{Ballester, M.A.G.}, \bibinfo{year}{2018}.
\newblock \bibinfo{title}{Fully automatic detection and segmentation of
  abdominal aortic thrombus in post-operative cta images using deep
  convolutional neural networks}.
\newblock \bibinfo{journal}{Medical Image Analysis} \bibinfo{volume}{46},
  \bibinfo{pages}{202--214}.
\newblock \DOIprefix\doi{10.1016/j.media.2018.03.010}.
\bibitem[{Martínez-Mera et~al.(2013)Martínez-Mera, Tahoces and
  Carreira}]{Mera_2013}
\bibinfo{author}{Martínez-Mera, J.A.}, \bibinfo{author}{Tahoces, P.G.},
  \bibinfo{author}{Carreira, J.M.}, \bibinfo{year}{2013}.
\newblock \bibinfo{title}{A hybrid method based on level set and \mbox{3D}
  region growing for segmentation of the thoracic aorta}.
\newblock \bibinfo{journal}{Computer Aided Surgery} \bibinfo{volume}{18(5-6)},
  \bibinfo{pages}{109--117}.
\bibitem[{Maton et~al.(1995)Maton, Hopkins, McLaughlin, Johnson, Warner, LaHart
  and Wright}]{Maton_1995}
\bibinfo{author}{Maton, A.}, \bibinfo{author}{Hopkins, J.},
  \bibinfo{author}{McLaughlin, C.W.}, \bibinfo{author}{Johnson, S.},
  \bibinfo{author}{Warner, M.Q.}, \bibinfo{author}{LaHart, D.},
  \bibinfo{author}{Wright, J.D.}, \bibinfo{year}{1995}.
\newblock \bibinfo{title}{Human Biology Health}.
\newblock \bibinfo{publisher}{Englewood Cliffs, New Jersey: Prentice Hall}.
\bibitem[{Mistelbauer et~al.(2016)Mistelbauer, Schmidt, Sailer, , Bäumler,
  Walters and Fleischmann}]{Mistelbauer_2016}
\bibinfo{author}{Mistelbauer, G.}, \bibinfo{author}{Schmidt, J.},
  \bibinfo{author}{Sailer, A.}, , \bibinfo{author}{Bäumler, K.},
  \bibinfo{author}{Walters, S.}, \bibinfo{author}{Fleischmann, D.},
  \bibinfo{year}{2016}.
\newblock \bibinfo{title}{Aortic dissection maps: comprehensive visualization
  of aortic dissections for risk assessment}, in: \bibinfo{booktitle}{VCBM '16:
  Proceedings of the Eurographics Workshop on Visual Computing for Biology and
  Medicine}, p. \bibinfo{pages}{143–152}.
\bibitem[{Moccia et~al.(2018)Moccia, di~Milano, Mattos, Momi and
  Hadji}]{Moccia_2018}
\bibinfo{author}{Moccia, S.}, \bibinfo{author}{di~Milano, E.D.M.P.},
  \bibinfo{author}{Mattos, L.S.}, \bibinfo{author}{Momi, E.D.},
  \bibinfo{author}{Hadji, S.E.}, \bibinfo{year}{2018}.
\newblock \bibinfo{title}{Blood vessel segmentation algorithms-review of
  methods, datasets and evaluation metrics}.
\newblock \URLprefix \url{https://www.researchgate.net/publication/331299499}.
\bibitem[{Morais et~al.(2017)Morais, Vilaça, Queirós, Bourier, Deisenhofer,
  Tavares and D'hooge}]{Morais_2017}
\bibinfo{author}{Morais, P.}, \bibinfo{author}{Vilaça, J.L.},
  \bibinfo{author}{Queirós, S.}, \bibinfo{author}{Bourier, F.},
  \bibinfo{author}{Deisenhofer, I.}, \bibinfo{author}{Tavares, J.M.R.},
  \bibinfo{author}{D'hooge, J.}, \bibinfo{year}{2017}.
\newblock \bibinfo{title}{A competitive strategy for atrial and aortic tract
  segmentation based on deformable models}.
\newblock \bibinfo{journal}{Medical Image Analysis} \bibinfo{volume}{42},
  \bibinfo{pages}{102--116}.
\newblock \DOIprefix\doi{10.1016/j.media.2017.07.007}.
\bibitem[{Morris et~al.(2019)Morris, Ghanem, Pantelic, Walker, Han and
  Glide-Hurst}]{Morris_2019}
\bibinfo{author}{Morris, E.D.}, \bibinfo{author}{Ghanem, A.I.},
  \bibinfo{author}{Pantelic, M.V.}, \bibinfo{author}{Walker, E.M.},
  \bibinfo{author}{Han, X.}, \bibinfo{author}{Glide-Hurst, C.K.},
  \bibinfo{year}{2019}.
\newblock \bibinfo{title}{Cardiac substructure segmentation and dosimetry using
  a novel hybrid magnetic resonance and computed tomography cardiac atlas}.
\newblock \bibinfo{journal}{International Journal of Radiation Oncology Biology
  Physics} \bibinfo{volume}{103}, \bibinfo{pages}{985--993}.
\newblock \DOIprefix\doi{10.1016/j.ijrobp.2018.11.025}.
\bibitem[{Nienaber(2013)}]{Nienaber_2013}
\bibinfo{author}{Nienaber, C.A.}, \bibinfo{year}{2013}.
\newblock \bibinfo{title}{The role of imaging in acute aortic syndromes}.
\newblock \bibinfo{journal}{European Heart Journal Cardiovascular Imaging}
  \bibinfo{volume}{14}, \bibinfo{pages}{15--23}.
\newblock \DOIprefix\doi{10.1093/ehjci/jes215}.
\bibitem[{Nienaber et~al.(2016)Nienaber, Clough, Sakalihasan, Suzuki, Gibbs,
  Mussa, Jenkins, Thompson, Evangelista, Yeh, Cheshire, Rosendahl and
  Pepper}]{Nienaber_2016ADRev}
\bibinfo{author}{Nienaber, C.A.}, \bibinfo{author}{Clough, R.E.},
  \bibinfo{author}{Sakalihasan, N.}, \bibinfo{author}{Suzuki, T.},
  \bibinfo{author}{Gibbs, R.}, \bibinfo{author}{Mussa, F.},
  \bibinfo{author}{Jenkins, M.P.}, \bibinfo{author}{Thompson, M.M.},
  \bibinfo{author}{Evangelista, A.}, \bibinfo{author}{Yeh, J.S.},
  \bibinfo{author}{Cheshire, N.}, \bibinfo{author}{Rosendahl, U.},
  \bibinfo{author}{Pepper, J.}, \bibinfo{year}{2016}.
\newblock \bibinfo{title}{Aortic dissection}.
\newblock \bibinfo{journal}{Nature Reviews Diesease Primers}
  \bibinfo{volume}{21}, \bibinfo{pages}{16053}.
\bibitem[{Olabarriaga et~al.(2005)Olabarriaga, Rouet, Fradkin, Breeuwer and
  Niessen}]{Olabarriaga_2005}
\bibinfo{author}{Olabarriaga, S.}, \bibinfo{author}{Rouet, j.m.},
  \bibinfo{author}{Fradkin, M.}, \bibinfo{author}{Breeuwer, M.},
  \bibinfo{author}{Niessen, W.}, \bibinfo{year}{2005}.
\newblock \bibinfo{title}{Segmentation of thrombus in abdominal aortic
  aneurysms from cta with non-parametric statistical grey level appearance
  modelling}.
\newblock \bibinfo{journal}{IEEE transactions on medical imaging}
  \bibinfo{volume}{24}, \bibinfo{pages}{477--85}.
\newblock \DOIprefix\doi{10.1109/TMI.2004.843260}.
\bibitem[{Osada et~al.(2018)Osada, Kyogoku, Matsuo and
  Kanemitsu}]{osada2018histopathological}
\bibinfo{author}{Osada, H.}, \bibinfo{author}{Kyogoku, M.},
  \bibinfo{author}{Matsuo, T.}, \bibinfo{author}{Kanemitsu, N.},
  \bibinfo{year}{2018}.
\newblock \bibinfo{title}{Histopathological evaluation of aortic dissection: a
  comparison of congenital versus acquired aortic wall weakness}.
\newblock \bibinfo{journal}{Interactive CardioVascular and Thoracic Surgery}
  \bibinfo{volume}{27}, \bibinfo{pages}{277--283}.
\bibitem[{Pepe et~al.(2020)Pepe, Li, Rolf-Pissarczyk, Gsaxner, Chen, Holzapfel
  and Egger}]{Pepe_2020}
\bibinfo{author}{Pepe, A.}, \bibinfo{author}{Li, J.},
  \bibinfo{author}{Rolf-Pissarczyk, M.}, \bibinfo{author}{Gsaxner, C.},
  \bibinfo{author}{Chen, X.}, \bibinfo{author}{Holzapfel, G.A.},
  \bibinfo{author}{Egger, J.}, \bibinfo{year}{2020}.
\newblock \bibinfo{title}{Detection, segmentation, simulation and visualization
  of aortic dissections: A review}.
\newblock \bibinfo{journal}{Medical Image Analysis} \bibinfo{volume}{65}.
\newblock \DOIprefix\doi{10.1016/j.media.2020.101773}.
\bibitem[{Petitjean et~al.(2019)Petitjean, Ruan, Lambert and
  Dubray}]{DBLP:conf/isbi/2019segthor}
\bibinfo{editor}{Petitjean, C.}, \bibinfo{editor}{Ruan, S.},
  \bibinfo{editor}{Lambert, Z.}, \bibinfo{editor}{Dubray, B.} (Eds.),
  \bibinfo{year}{2019}.
\newblock \bibinfo{title}{Proceedings of the 2019 Challenge on Segmentation of
  THoracic Organs at Risk in {CT} Images, SegTHOR@ISBI 2019, April 8, 2019}.
  volume \bibinfo{volume}{2349} of \textit{\bibinfo{series}{{CEUR} Workshop
  Proceedings}}, \bibinfo{publisher}{CEUR-WS.org}.
\newblock \URLprefix \url{http://ceur-ws.org/Vol-2349}.
\bibitem[{Pinheiro et~al.(2016)Pinheiro, Lin, Collobert and
  Doll{\'a}r}]{Pinheiro_2016}
\bibinfo{author}{Pinheiro, P.O.}, \bibinfo{author}{Lin, T.Y.},
  \bibinfo{author}{Collobert, R.}, \bibinfo{author}{Doll{\'a}r, P.},
  \bibinfo{year}{2016}.
\newblock \bibinfo{title}{Learning to refine object segments}, in:
  \bibinfo{editor}{Leibe, B.}, \bibinfo{editor}{Matas, J.},
  \bibinfo{editor}{Sebe, N.}, \bibinfo{editor}{Welling, M.} (Eds.),
  \bibinfo{booktitle}{Computer Vision -- ECCV 2016},
  \bibinfo{publisher}{Springer International Publishing},
  \bibinfo{address}{Cham}. pp. \bibinfo{pages}{75--91}.
\bibitem[{Pock et~al.(2005)Pock, Beichel and Bischof}]{Pock_2005}
\bibinfo{author}{Pock, T.}, \bibinfo{author}{Beichel, R.},
  \bibinfo{author}{Bischof, H.}, \bibinfo{year}{2005}.
\newblock \bibinfo{title}{A novel robust tube detection filter for 3d
  centerline extraction}, in: \bibinfo{booktitle}{Scandinavian Conference on
  Image Analysis}, \bibinfo{organization}{Springer}. pp.
  \bibinfo{pages}{481--490}.
\bibitem[{Radl et~al.(2022a)Radl, Jin, Pepe, Li, Gsaxner, hua Zhao and
  Egger}]{Radl2022avtData}
\bibinfo{author}{Radl, L.}, \bibinfo{author}{Jin, Y.}, \bibinfo{author}{Pepe,
  A.}, \bibinfo{author}{Li, J.}, \bibinfo{author}{Gsaxner, C.},
  \bibinfo{author}{hua Zhao, F.}, \bibinfo{author}{Egger, J.},
  \bibinfo{year}{2022}a.
\newblock \bibinfo{title}{{Aortic Vessel Tree (AVT) CTA Datasets and
  Segmentations}} \URLprefix
  \url{https://figshare.com/articles/dataset/Aortic_Vessel_Tree_AVT_CTA_Datasets_and_Segmentations/14806362},
  \DOIprefix\doi{10.6084/m9.figshare.14806362.v1}.
\bibitem[{Radl et~al.(2022b)Radl, Jin, Pepe, Li, Gsaxner, Zhao and
  Egger}]{radl2022avtPaper}
\bibinfo{author}{Radl, L.}, \bibinfo{author}{Jin, Y.}, \bibinfo{author}{Pepe,
  A.}, \bibinfo{author}{Li, J.}, \bibinfo{author}{Gsaxner, C.},
  \bibinfo{author}{Zhao, F.h.}, \bibinfo{author}{Egger, J.},
  \bibinfo{year}{2022}b.
\newblock \bibinfo{title}{Avt: Multicenter aortic vessel tree cta dataset
  collection with ground truth segmentation masks}.
\newblock \bibinfo{journal}{Data in Brief} , \bibinfo{pages}{107801}.
\bibitem[{Ritter et~al.(2006)Ritter, Hansen, Dicken, Konrad, Preim and
  Peitgen}]{ritter2006real}
\bibinfo{author}{Ritter, F.}, \bibinfo{author}{Hansen, C.},
  \bibinfo{author}{Dicken, V.}, \bibinfo{author}{Konrad, O.},
  \bibinfo{author}{Preim, B.}, \bibinfo{author}{Peitgen, H.O.},
  \bibinfo{year}{2006}.
\newblock \bibinfo{title}{Real-time illustration of vascular structures}.
\newblock \bibinfo{journal}{IEEE Transactions on Visualization and Computer
  Graphics} \bibinfo{volume}{12}, \bibinfo{pages}{877--884}.
\bibitem[{Robben et~al.(2014)Robben, Türetken, Sunaert, Thijs, Wilms, Fua,
  Maes and Suetens}]{Robben_2014}
\bibinfo{author}{Robben, D.}, \bibinfo{author}{Türetken, E.},
  \bibinfo{author}{Sunaert, S.}, \bibinfo{author}{Thijs, V.},
  \bibinfo{author}{Wilms, G.}, \bibinfo{author}{Fua, P.},
  \bibinfo{author}{Maes, F.}, \bibinfo{author}{Suetens, P.},
  \bibinfo{year}{2014}.
\newblock \bibinfo{title}{Lncs 8673 - simultaneous segmentation and anatomical
  labeling of the cerebral vasculature}.
\bibitem[{Rogers et~al.(2011)Rogers, Hermann, Booher, Nienaber, Williams,
  Kazerooni, Froehlich, O'Gara, Montgomery, Cooper, Harris, Hutchison,
  Evangelista, Isselbacher and Eagle}]{Rogers_2011}
\bibinfo{author}{Rogers, A.M.}, \bibinfo{author}{Hermann, L.K.},
  \bibinfo{author}{Booher, A.M.}, \bibinfo{author}{Nienaber, C.A.},
  \bibinfo{author}{Williams, D.M.}, \bibinfo{author}{Kazerooni, E.A.},
  \bibinfo{author}{Froehlich, J.B.}, \bibinfo{author}{O'Gara, P.T.},
  \bibinfo{author}{Montgomery, D.G.}, \bibinfo{author}{Cooper, J.V.},
  \bibinfo{author}{Harris, K.M.}, \bibinfo{author}{Hutchison, S.},
  \bibinfo{author}{Evangelista, A.}, \bibinfo{author}{Isselbacher, E.M.},
  \bibinfo{author}{Eagle, K.A.}, \bibinfo{year}{2011}.
\newblock \bibinfo{title}{Sensitivity of the aortic dissection detection risk
  score, a novel guideline-based tool for identification of acute aortic
  dissection at initial presentation: Results from the international registry
  of acute aortic dissection}.
\newblock \bibinfo{journal}{Circulation} \bibinfo{volume}{123},
  \bibinfo{pages}{2213--2218}.
\newblock \DOIprefix\doi{10.1161/CIRCULATIONAHA.110.988568}.
\bibitem[{Ronneberger et~al.(2015)Ronneberger, Fischer and
  Brox}]{Ronneberger_2015}
\bibinfo{author}{Ronneberger, O.}, \bibinfo{author}{Fischer, P.},
  \bibinfo{author}{Brox, T.}, \bibinfo{year}{2015}.
\newblock \bibinfo{title}{U-net: Convolutional networks for biomedical image
  segmentation}.
\newblock \bibinfo{journal}{CoRR} \bibinfo{volume}{abs/1505.04597}.
\newblock \URLprefix \url{http://arxiv.org/abs/1505.04597}.
\bibitem[{{Rueckert} et~al.(1997){Rueckert}, {Burger}, {Forbat}, {Mohiaddin}
  and {Yang}}]{Rueckert_1997}
\bibinfo{author}{{Rueckert}, D.}, \bibinfo{author}{{Burger}, P.},
  \bibinfo{author}{{Forbat}, S.M.}, \bibinfo{author}{{Mohiaddin}, R.D.},
  \bibinfo{author}{{Yang}, G.Z.}, \bibinfo{year}{1997}.
\newblock \bibinfo{title}{Automatic tracking of the aorta in cardiovascular mr
  images using deformable models}.
\newblock \bibinfo{journal}{IEEE Transactions on Medical Imaging}
  \bibinfo{volume}{16}, \bibinfo{pages}{581--590}.
\newblock \DOIprefix\doi{10.1109/42.640747}.
\bibitem[{Schmied et~al.(2021)Schmied, Pepe and Egger}]{schmied2021patch}
\bibinfo{author}{Schmied, M.}, \bibinfo{author}{Pepe, A.},
  \bibinfo{author}{Egger, J.}, \bibinfo{year}{2021}.
\newblock \bibinfo{title}{A patch-based-approach for aortic landmarking}, in:
  \bibinfo{booktitle}{Medical Imaging 2021: Biomedical Applications in
  Molecular, Structural, and Functional Imaging},
  \bibinfo{organization}{International Society for Optics and Photonics}. p.
  \bibinfo{pages}{1160010}.
\bibitem[{Selver and Kavur(2016)}]{Selver_2016}
\bibinfo{author}{Selver, M.A.}, \bibinfo{author}{Kavur, A.E.},
  \bibinfo{year}{2016}.
\newblock \bibinfo{title}{Implementation and use of 3d pairwise geodesic
  distance fields for seeding abdominal aortic vessels}.
\newblock \bibinfo{journal}{International Journal of Computer Assisted
  Radiology and Surgery} \bibinfo{volume}{11}, \bibinfo{pages}{803--816}.
\newblock \DOIprefix\doi{10.1007/s11548-015-1321-z}.
\bibitem[{Sethian and Vladimirsky(2000)}]{Sethian_2000}
\bibinfo{author}{Sethian, J.A.}, \bibinfo{author}{Vladimirsky, A.},
  \bibinfo{year}{2000}.
\newblock \bibinfo{title}{Fast methods for the eikonal and related
  hamilton{\textendash} jacobi equations on unstructured meshes}.
\newblock \bibinfo{journal}{Proceedings of the National Academy of Sciences}
  \bibinfo{volume}{97}, \bibinfo{pages}{5699--5703}.
\newblock \URLprefix \url{https://www.pnas.org/content/97/11/5699},
  \DOIprefix\doi{10.1073/pnas.090060097},
  \href{http://arxiv.org/abs/https://www.pnas.org/content/97/11/5699.full.pdf}{\tt
  arXiv:https://www.pnas.org/content/97/11/5699.full.pdf}.
\bibitem[{Shahzad et~al.(2015)Shahzad, Dzyubachyk, Staring, Kullberg,
  Johansson, Ahlström, Lelieveldt and {van der Geest}}]{Shahzad_2015}
\bibinfo{author}{Shahzad, R.}, \bibinfo{author}{Dzyubachyk, O.},
  \bibinfo{author}{Staring, M.}, \bibinfo{author}{Kullberg, J.},
  \bibinfo{author}{Johansson, L.}, \bibinfo{author}{Ahlström, H.},
  \bibinfo{author}{Lelieveldt, B.P.}, \bibinfo{author}{{van der Geest}, R.J.},
  \bibinfo{year}{2015}.
\newblock \bibinfo{title}{Automated extraction and labelling of the arterial
  tree from whole-body mra data}.
\newblock \bibinfo{journal}{Medical Image Analysis} \bibinfo{volume}{24},
  \bibinfo{pages}{28--40}.
\newblock \URLprefix
  \url{https://www.sciencedirect.com/science/article/pii/S1361841515000754},
  \DOIprefix\doi{https://doi.org/10.1016/j.media.2015.05.008}.
\bibitem[{Shamseer et~al.(2015)Shamseer, Moher, Clarke, Ghersi, Liberati,
  Petticrew, Shekelle and Stewart}]{Shamseerg_2015}
\bibinfo{author}{Shamseer, L.}, \bibinfo{author}{Moher, D.},
  \bibinfo{author}{Clarke, M.}, \bibinfo{author}{Ghersi, D.},
  \bibinfo{author}{Liberati, A.}, \bibinfo{author}{Petticrew, M.},
  \bibinfo{author}{Shekelle, P.}, \bibinfo{author}{Stewart, L.A.},
  \bibinfo{year}{2015}.
\newblock \bibinfo{title}{Preferred reporting items for systematic review and
  meta-analysis protocols (prisma-p) 2015: elaboration and explanation}.
\newblock \bibinfo{journal}{BMJ} \bibinfo{volume}{349}.
\newblock \URLprefix \url{https://www.bmj.com/content/349/bmj.g7647},
  \DOIprefix\doi{10.1136/bmj.g7647},
  \href{http://arxiv.org/abs/https://www.bmj.com/content/349/bmj.g7647.full.pdf}{\tt
  arXiv:https://www.bmj.com/content/349/bmj.g7647.full.pdf}.
\bibitem[{Shen et~al.(2017)Shen, Wu and Suk}]{shen2017deep}
\bibinfo{author}{Shen, D.}, \bibinfo{author}{Wu, G.}, \bibinfo{author}{Suk,
  H.I.}, \bibinfo{year}{2017}.
\newblock \bibinfo{title}{Deep learning in medical image analysis}.
\newblock \bibinfo{journal}{Annual review of biomedical engineering}
  \bibinfo{volume}{19}, \bibinfo{pages}{221--248}.
\bibitem[{Sieren et~al.(2022)Sieren, Widmann, Weiss, Moltz, Link, Wegner,
  Stahlberg, Horn, Oecherting, Goltz, Barkhausen and
  Frydrychowicz}]{Sieren_2022}
\bibinfo{author}{Sieren, M.M.}, \bibinfo{author}{Widmann, C.},
  \bibinfo{author}{Weiss, N.}, \bibinfo{author}{Moltz, J.H.},
  \bibinfo{author}{Link, F.}, \bibinfo{author}{Wegner, F.},
  \bibinfo{author}{Stahlberg, E.}, \bibinfo{author}{Horn, M.},
  \bibinfo{author}{Oecherting, T.H.}, \bibinfo{author}{Goltz, J.P.},
  \bibinfo{author}{Barkhausen, J.}, \bibinfo{author}{Frydrychowicz, A.},
  \bibinfo{year}{2022}.
\newblock \bibinfo{title}{Automated segmentation and quantification of the
  healthy and diseased aorta in ct angiographies using a dedicated deep
  learning approach}.
\newblock \bibinfo{journal}{Eur Radiol} \bibinfo{volume}{32},
  \bibinfo{pages}{690--701}.
\newblock \DOIprefix\doi{10.1007/s00330-021-08130-2}.
\bibitem[{Simpson et~al.(2019)Simpson, Antonelli, Bakas, Bilello, Farahani, van
  Ginneken, Kopp-Schneider, Landman, Litjens, Menze, Ronneberger, Summers,
  Bilic, Christ, Do, Gollub, Golia-Pernicka, Heckers, Jarnagin, McHugo, Napel,
  Vorontsov, Maier-Hein and Cardoso}]{Simpson_2019}
\bibinfo{author}{Simpson, A.L.}, \bibinfo{author}{Antonelli, M.},
  \bibinfo{author}{Bakas, S.}, \bibinfo{author}{Bilello, M.},
  \bibinfo{author}{Farahani, K.}, \bibinfo{author}{van Ginneken, B.},
  \bibinfo{author}{Kopp-Schneider, A.}, \bibinfo{author}{Landman, B.A.},
  \bibinfo{author}{Litjens, G.}, \bibinfo{author}{Menze, B.},
  \bibinfo{author}{Ronneberger, O.}, \bibinfo{author}{Summers, R.M.},
  \bibinfo{author}{Bilic, P.}, \bibinfo{author}{Christ, P.F.},
  \bibinfo{author}{Do, R.K.G.}, \bibinfo{author}{Gollub, M.},
  \bibinfo{author}{Golia-Pernicka, J.}, \bibinfo{author}{Heckers, S.H.},
  \bibinfo{author}{Jarnagin, W.R.}, \bibinfo{author}{McHugo, M.K.},
  \bibinfo{author}{Napel, S.}, \bibinfo{author}{Vorontsov, E.},
  \bibinfo{author}{Maier-Hein, L.}, \bibinfo{author}{Cardoso, M.J.},
  \bibinfo{year}{2019}.
\newblock \bibinfo{title}{A large annotated medical image dataset for the
  development and evaluation of segmentation algorithms}.
\newblock \href{http://arxiv.org/abs/1902.09063}{\tt arXiv:1902.09063}.
\bibitem[{Sreejini and Govindan(2015)}]{Sreejini_2015}
\bibinfo{author}{Sreejini, K.S.}, \bibinfo{author}{Govindan, V.K.},
  \bibinfo{year}{2015}.
\newblock \bibinfo{title}{Improved multiscale matched filter for retina vessel
  segmentation using pso algorithm}.
\newblock \bibinfo{journal}{Egyptian Informatics Journal} \bibinfo{volume}{16},
  \bibinfo{pages}{253--260}.
\newblock \DOIprefix\doi{10.1016/j.eij.2015.06.004}.
\bibitem[{Subasic et~al.(2000)Subasic, Loncaric and Sorantin}]{Subasic_2000}
\bibinfo{author}{Subasic, M.}, \bibinfo{author}{Loncaric, S.},
  \bibinfo{author}{Sorantin, E.}, \bibinfo{year}{2000}.
\newblock \bibinfo{title}{3-d image analysis of abdominal aortic aneurysm}, in:
  \bibinfo{booktitle}{Medical Infobahn for Europe}. \bibinfo{publisher}{IOS
  Press}, pp. \bibinfo{pages}{1195--1200}.
\bibitem[{Subramanyan et~al.(2003)Subramanyan, Steinmiller, Sifri and
  Boll}]{Subramanyan_2003}
\bibinfo{author}{Subramanyan, K.}, \bibinfo{author}{Steinmiller, M.},
  \bibinfo{author}{Sifri, D.}, \bibinfo{author}{Boll, D.},
  \bibinfo{year}{2003}.
\newblock \bibinfo{title}{{Automatic aortic vessel tree extraction and thrombus
  detection in multislice CT}}, in: \bibinfo{editor}{Sonka, M.},
  \bibinfo{editor}{Fitzpatrick, J.M.} (Eds.), \bibinfo{booktitle}{Medical
  Imaging 2003: Image Processing}, \bibinfo{organization}{International Society
  for Optics and Photonics}. \bibinfo{publisher}{SPIE}. pp.
  \bibinfo{pages}{1629 -- 1638}.
\newblock \URLprefix \url{https://doi.org/10.1117/12.483542}.
\bibitem[{Taha and Hanbury(2015)}]{taha2015metrics}
\bibinfo{author}{Taha, A.A.}, \bibinfo{author}{Hanbury, A.},
  \bibinfo{year}{2015}.
\newblock \bibinfo{title}{Metrics for evaluating 3d medical image segmentation:
  analysis, selection, and tool}.
\newblock \bibinfo{journal}{BMC medical imaging} \bibinfo{volume}{15},
  \bibinfo{pages}{1--28}.
\bibitem[{Tahoces et~al.(2019)Tahoces, Alvarez, González, Cuenca, Trujillo,
  Santana-Cedrés, Esclarín, Gomez, Mazorra, Alemán-Flores and
  Carreira}]{Tahoces_2019}
\bibinfo{author}{Tahoces, P.G.}, \bibinfo{author}{Alvarez, L.},
  \bibinfo{author}{González, E.}, \bibinfo{author}{Cuenca, C.},
  \bibinfo{author}{Trujillo, A.}, \bibinfo{author}{Santana-Cedrés, D.},
  \bibinfo{author}{Esclarín, J.}, \bibinfo{author}{Gomez, L.},
  \bibinfo{author}{Mazorra, L.}, \bibinfo{author}{Alemán-Flores, M.},
  \bibinfo{author}{Carreira, J.M.}, \bibinfo{year}{2019}.
\newblock \bibinfo{title}{Automatic estimation of the aortic lumen geometry by
  ellipse tracking}.
\newblock \bibinfo{journal}{International Journal of Computer Assisted
  Radiology and Surgery} \bibinfo{volume}{14}, \bibinfo{pages}{345--355}.
\newblock \DOIprefix\doi{10.1007/s11548-018-1861-0}.
\bibitem[{Tang et~al.(2012)Tang, van Walsum, van Onkelen, Hameeteman, Klein,
  Schaap, Tori, van~den Bouwhuijsen, Witteman, van~der Lugt, van Vliet and
  Niessen}]{Tang_2012}
\bibinfo{author}{Tang, H.}, \bibinfo{author}{van Walsum, T.},
  \bibinfo{author}{van Onkelen, R.S.}, \bibinfo{author}{Hameeteman, R.},
  \bibinfo{author}{Klein, S.}, \bibinfo{author}{Schaap, M.},
  \bibinfo{author}{Tori, F.L.}, \bibinfo{author}{van~den Bouwhuijsen, Q.J.},
  \bibinfo{author}{Witteman, J.C.}, \bibinfo{author}{van~der Lugt, A.},
  \bibinfo{author}{van Vliet, L.J.}, \bibinfo{author}{Niessen, W.J.},
  \bibinfo{year}{2012}.
\newblock \bibinfo{title}{Semiautomatic carotid lumen segmentation for
  quantification of lumen geometry in multispectral mri}.
\newblock \bibinfo{journal}{Medical Image Analysis} \bibinfo{volume}{16},
  \bibinfo{pages}{1202--1215}.
\newblock \DOIprefix\doi{10.1016/j.media.2012.05.014}.
\bibitem[{{Terzopoulos} and {Metaxas}(1991)}]{Terzopoulos_1991}
\bibinfo{author}{{Terzopoulos}, D.}, \bibinfo{author}{{Metaxas}, D.},
  \bibinfo{year}{1991}.
\newblock \bibinfo{title}{Dynamic 3d models with local and global deformations:
  deformable superquadrics}.
\newblock \bibinfo{journal}{IEEE Transactions on Pattern Analysis and Machine
  Intelligence} \bibinfo{volume}{13}, \bibinfo{pages}{703--714}.
\newblock \DOIprefix\doi{10.1109/34.85659}.
\bibitem[{Tortora and Nielsen(2016)}]{Tortora_2016}
\bibinfo{author}{Tortora, G.J.}, \bibinfo{author}{Nielsen, M.T.},
  \bibinfo{year}{2016}.
\newblock \bibinfo{title}{Principles of Human Anatomy, 14th Edition}.
\newblock \bibinfo{publisher}{WILEY}.
\bibitem[{Trullo et~al.(2017)Trullo, Petitjean, Ruan, Dubray, Nie and
  Shen}]{Trullo_2017}
\bibinfo{author}{Trullo, R.}, \bibinfo{author}{Petitjean, C.},
  \bibinfo{author}{Ruan, S.}, \bibinfo{author}{Dubray, B.},
  \bibinfo{author}{Nie, D.}, \bibinfo{author}{Shen, D.}, \bibinfo{year}{2017}.
\newblock \bibinfo{title}{Segmentation of organs at risk in thoracic ct images
  using a sharpmask architecture and conditional random fields},
  \bibinfo{publisher}{IEEE Computer Society}. pp. \bibinfo{pages}{1003--1006}.
\newblock \DOIprefix\doi{10.1109/ISBI.2017.7950685}.
\bibitem[{Tsai et~al.(2005)Tsai, Nienaber and Eagle}]{tsai2005acute}
\bibinfo{author}{Tsai, T.T.}, \bibinfo{author}{Nienaber, C.A.},
  \bibinfo{author}{Eagle, K.A.}, \bibinfo{year}{2005}.
\newblock \bibinfo{title}{Acute aortic syndromes}.
\newblock \bibinfo{journal}{Circulation} \bibinfo{volume}{112},
  \bibinfo{pages}{3802--3813}.
\bibitem[{Volonghi et~al.(2016)Volonghi, Tresoldi, Cadioli, Usuelli, Ponzini,
  Morbiducci, Esposito and Rizzo}]{Volonghi_2016}
\bibinfo{author}{Volonghi, P.}, \bibinfo{author}{Tresoldi, D.},
  \bibinfo{author}{Cadioli, M.}, \bibinfo{author}{Usuelli, A.M.},
  \bibinfo{author}{Ponzini, R.}, \bibinfo{author}{Morbiducci, U.},
  \bibinfo{author}{Esposito, A.}, \bibinfo{author}{Rizzo, G.},
  \bibinfo{year}{2016}.
\newblock \bibinfo{title}{Automatic extraction of three-dimensional thoracic
  aorta geometric model from phase contrast mri for morphometric and
  hemodynamic characterization}.
\newblock \bibinfo{journal}{Magnetic Resonance in Medicine}
  \bibinfo{volume}{75}, \bibinfo{pages}{873--882}.
\newblock \DOIprefix\doi{10.1002/mrm.25630}.
\bibitem[{Wang et~al.(2012)Wang, Heimann, Lo, Sumkauskaite, Puderbach,
  de~Bruijne, Meinzer and Wegner}]{Wang_2012}
\bibinfo{author}{Wang, X.}, \bibinfo{author}{Heimann, T.}, \bibinfo{author}{Lo,
  P.}, \bibinfo{author}{Sumkauskaite, M.}, \bibinfo{author}{Puderbach, M.},
  \bibinfo{author}{de~Bruijne, M.}, \bibinfo{author}{Meinzer, H.P.},
  \bibinfo{author}{Wegner, I.}, \bibinfo{year}{2012}.
\newblock \bibinfo{title}{Statistical tracking of tree-like tubular structures
  with efficient branching detection in 3d medical image data.}
\newblock \bibinfo{journal}{Physics in medicine and biology}
  \bibinfo{volume}{57}, \bibinfo{pages}{5325--5342}.
\newblock \DOIprefix\doi{10.1088/0031-9155/57/16/5325}.
\bibitem[{Wang et~al.(2017)Wang, Seguro, Kao, Zhang, Faraji, Zhu, Haraldsson,
  Hope, Saloner and Liu}]{Wang_2017}
\bibinfo{author}{Wang, Y.}, \bibinfo{author}{Seguro, F.}, \bibinfo{author}{Kao,
  E.}, \bibinfo{author}{Zhang, Y.}, \bibinfo{author}{Faraji, F.},
  \bibinfo{author}{Zhu, C.}, \bibinfo{author}{Haraldsson, H.},
  \bibinfo{author}{Hope, M.}, \bibinfo{author}{Saloner, D.},
  \bibinfo{author}{Liu, J.}, \bibinfo{year}{2017}.
\newblock \bibinfo{title}{Segmentation of lumen and outer wall of abdominal
  aortic aneurysms from 3d black-blood mri with a registration based geodesic
  active contour model}.
\newblock \bibinfo{journal}{Medical Image Analysis} \bibinfo{volume}{40},
  \bibinfo{pages}{1--10}.
\newblock \DOIprefix\doi{10.1016/j.media.2017.05.005}.
\bibitem[{Wu et~al.(2013)Wu, Shen, Russell, Coselli and
  LeMaire}]{wu2013molecular}
\bibinfo{author}{Wu, D.}, \bibinfo{author}{Shen, Y.H.},
  \bibinfo{author}{Russell, L.}, \bibinfo{author}{Coselli, J.S.},
  \bibinfo{author}{LeMaire, S.A.}, \bibinfo{year}{2013}.
\newblock \bibinfo{title}{Molecular mechanisms of thoracic aortic dissection}.
\newblock \bibinfo{journal}{Journal of Surgical Research}
  \bibinfo{volume}{184}, \bibinfo{pages}{907--924}.
\bibitem[{Xie et~al.(2014)Xie, Padgett, Biancardi and Reeves}]{Xie_2014}
\bibinfo{author}{Xie, Y.}, \bibinfo{author}{Padgett, J.},
  \bibinfo{author}{Biancardi, A.M.}, \bibinfo{author}{Reeves, A.P.},
  \bibinfo{year}{2014}.
\newblock \bibinfo{title}{Automated aorta segmentation in low-dose chest ct
  images}.
\newblock \bibinfo{journal}{International Journal of Computer Assisted
  Radiology and Surgery} \bibinfo{volume}{9}, \bibinfo{pages}{211--219}.
\newblock \DOIprefix\doi{10.1007/s11548-013-0924-5}.
\bibitem[{Yu et~al.(2021)Yu, Gao, Wei, Liao, Xiao, Zhang, Yin and Lu}]{Yu_2021}
\bibinfo{author}{Yu, Y.}, \bibinfo{author}{Gao, Y.}, \bibinfo{author}{Wei, J.},
  \bibinfo{author}{Liao, F.}, \bibinfo{author}{Xiao, Q.},
  \bibinfo{author}{Zhang, J.}, \bibinfo{author}{Yin, W.}, \bibinfo{author}{Lu,
  B.}, \bibinfo{year}{2021}.
\newblock \bibinfo{title}{A three-dimensional deep convolutional neural network
  for automatic segmentation and diameter measurement of type b aortic
  dissection}.
\newblock \bibinfo{journal}{Korean journal of radiology} \bibinfo{volume}{22},
  \bibinfo{pages}{168--178}.
\newblock \DOIprefix\doi{10.3348/kjr.2020.0313}.
\bibitem[{Zhao et~al.(2022)Zhao, Zhao, Pang and Feng}]{Zhao_2022}
\bibinfo{author}{Zhao, J.}, \bibinfo{author}{Zhao, J.}, \bibinfo{author}{Pang,
  S.}, \bibinfo{author}{Feng, Q.}, \bibinfo{year}{2022}.
\newblock \bibinfo{title}{Segmentation of the true lumen of aorta dissection
  via morphology-constrained stepwise deep mesh regression}.
\newblock \bibinfo{journal}{IEEE Transactions on Medical Imaging}
  \bibinfo{volume}{41}, \bibinfo{pages}{1826--1836}.
\newblock \DOIprefix\doi{10.1109/TMI.2022.3150005}.
\bibitem[{Zheng et~al.(2010)Zheng, John, Liao, Boese, Kirschstein, Georgescu,
  Zhou, Kempfert, Walther, Brockmann and Comaniciu}]{Zheng_2010}
\bibinfo{author}{Zheng, Y.}, \bibinfo{author}{John, M.}, \bibinfo{author}{Liao,
  R.}, \bibinfo{author}{Boese, J.}, \bibinfo{author}{Kirschstein, U.},
  \bibinfo{author}{Georgescu, B.}, \bibinfo{author}{Zhou, S.K.},
  \bibinfo{author}{Kempfert, J.}, \bibinfo{author}{Walther, T.},
  \bibinfo{author}{Brockmann, G.}, \bibinfo{author}{Comaniciu, D.},
  \bibinfo{year}{2010}.
\newblock \bibinfo{title}{Automatic aorta segmentation and valve landmark
  detection in c-arm ct: Application to aortic valve implantation}, pp.
  \bibinfo{pages}{476--483}.
\newblock \DOIprefix\doi{10.1007/978-3-642-15705-9_58}.
\bibitem[{Zhong et~al.(2021)Zhong, Bian, Hatt and Burris}]{Zhong_2021}
\bibinfo{author}{Zhong, J.}, \bibinfo{author}{Bian, Z.}, \bibinfo{author}{Hatt,
  C.R.}, \bibinfo{author}{Burris, N.S.}, \bibinfo{year}{2021}.
\newblock \bibinfo{title}{{Segmentation of the thoracic aorta using an
  attention-gated u-net}}, in: \bibinfo{editor}{Mazurowski, M.A.},
  \bibinfo{editor}{Drukker, K.} (Eds.), \bibinfo{booktitle}{Medical Imaging
  2021: Computer-Aided Diagnosis}, \bibinfo{organization}{International Society
  for Optics and Photonics}. \bibinfo{publisher}{SPIE}. pp. \bibinfo{pages}{147
  -- 153}.
\newblock \URLprefix \url{https://doi.org/10.1117/12.2581947},
  \DOIprefix\doi{10.1117/12.2581947}.
\bibitem[{Zhuge et~al.(2006)Zhuge, Rubin, Sun and Napel}]{zhuge_2006}
\bibinfo{author}{Zhuge, F.}, \bibinfo{author}{Rubin, G.}, \bibinfo{author}{Sun,
  S.}, \bibinfo{author}{Napel, S.}, \bibinfo{year}{2006}.
\newblock \bibinfo{title}{An abdominal aortic aneurysm segmentation method:
  Level set with region and statistical information}.
\newblock \bibinfo{journal}{Medical physics} \bibinfo{volume}{33},
  \bibinfo{pages}{1440--53}.
\newblock \DOIprefix\doi{10.1118/1.2193247}.

\end{thebibliography}

%

\section*{ }
\includepdf[pages=-]{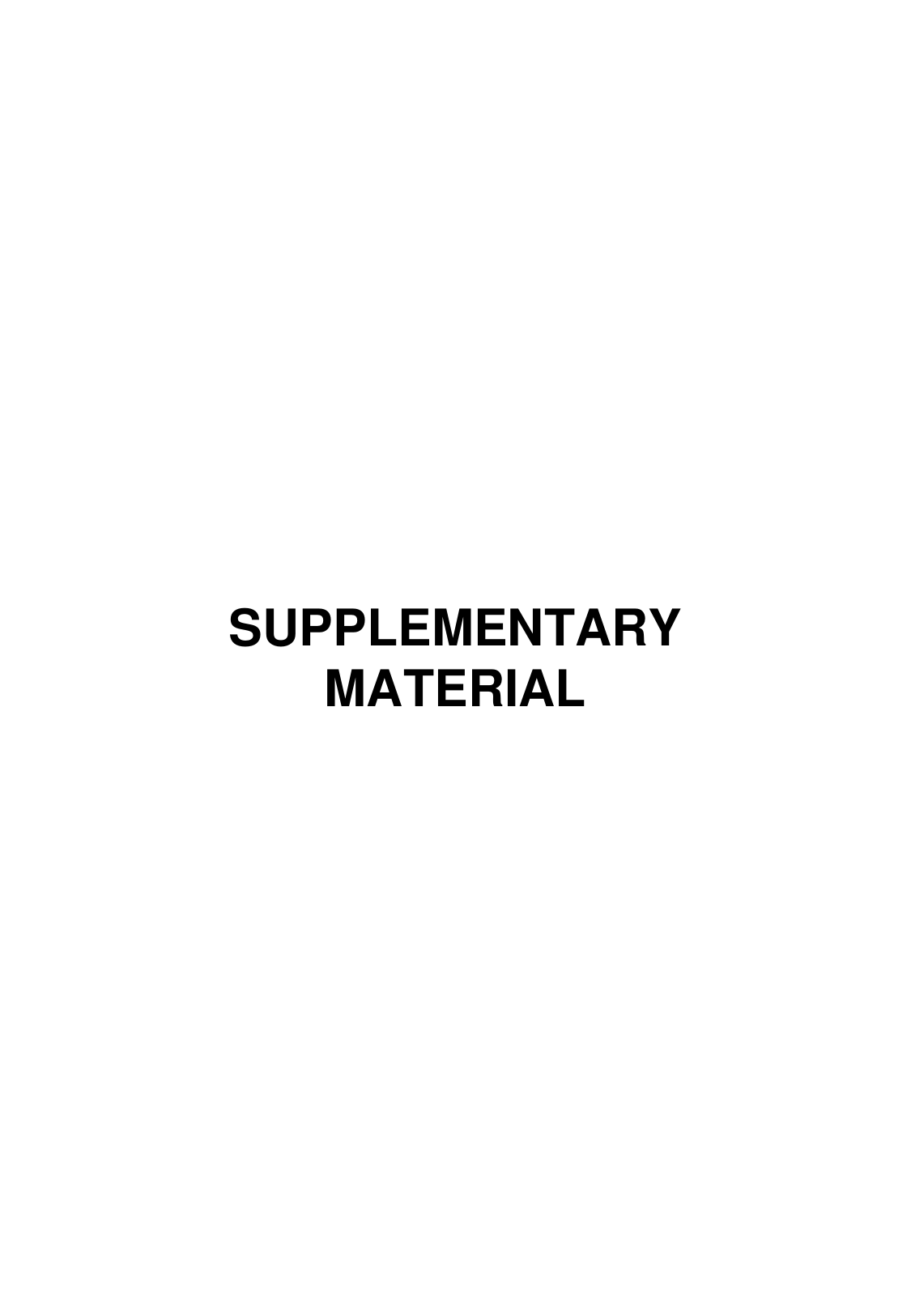}

\end{document}